\documentclass[aps,prd,amsmath,floats,floatfix,
  superscriptaddress,
  nofootinbib,
  notitlepage,
  10pt,
  showpacs]{revtex4-1}
  
\usepackage{graphicx,color}
\usepackage{amsmath,amssymb}
\usepackage{verbatim}
\usepackage{float}
\usepackage{dsfont}
\usepackage{amsfonts}
\usepackage{mathrsfs}
\usepackage{multirow}
\usepackage{times}
\usepackage{bm}
\usepackage{hyperref}
\usepackage{scalefnt}
\usepackage{xspace}
\usepackage{mathtools}
\usepackage{lipsum} 
\usepackage{physics}
\hypersetup{
  colorlinks=true,        
  linkcolor=blue,         
  citecolor=cyan,         
}
\usepackage{blindtext, rotating}
\usepackage{pifont}
\usepackage[normalem]{ulem}   
\usepackage{grffile}
\usepackage{booktabs}
\graphicspath{{./}}  

\usepackage{subfigure} %

\usepackage{booktabs, multirow} 
\usepackage{soul}
\usepackage{changepage,threeparttable} 
\usepackage{soul}
\newcommand{\re}{\mbox{Re}}

\def\Lie{\mathcal{L}}
\def\p{\partial}
\def\A{\mathcal{X}}
\def\Aphi{\A_{\phi}}

\begin{document}



\title{Full 3D nonlinear dynamics of charged and magnetized boson stars}
\author{V\'ictor Jaramillo }
\affiliation{Department of Astronomy, University of Science and Technology of China, Hefei, Anhui 230026, China}
\author{Dar\'io N\'u\~nez }
\affiliation{Instituto de Ciencias Nucleares, Universidad Nacional
  Aut\'onoma de M\'exico, Circuito Exterior C.U., A.P. 70-543,
  Coyoac\'an, M\'exico 04510, CdMx, M\'exico}
\affiliation{Centre for Research and Development in Mathematics and Applications (CIDMA),\\ 
Department of Mathematics, University of Aveiro, 3810-193 Aveiro, Portugal}
\author{Milton Ruiz}
\affiliation{Departament d'Astronomia i Astrof\'isica, Universitat de València, C/ Dr Moliner 50, 46100, Burjassot (València), Spain}
\author{Miguel Zilh\~ao}
\affiliation{Centre for Research and Development in Mathematics and Applications (CIDMA),\\ 
Department of Mathematics, University of Aveiro, 3810-193 Aveiro, Portugal}
%
\begin{abstract}
Gauged boson stars are exotic compact objects that can potentially mimic black holes or magnetized neutron stars in both their gravitational and electromagnetic signatures, offering a compelling new description or even an alternative explanation for various multimessenger phenomena. As a crucial step toward establishing boson stars as viable multimessenger sources, we perform 3D numerical simulations of the fully nonlinear Einstein-Maxwell-Klein-Gordon system, focusing on both spherical and axisymmetric boson star configurations that vary in their electromagnetic coupling between the neutral case up to values close to the critical case, and so their magnetic field content. For spherical configurations, we consistently find stable solutions. In contrast, for axially symmetric, electrically neutral, magnetized configurations, the dynamics are highly sensitive to the electromagnetic coupling. Configurations with stronger coupling develop a one-armed mode instability, which leads to collapse into black holes. Configurations with weaker coupling undergo a two-stage process: an initial bar-mode instability that triggers a one-armed spiral deformation. This eventually also results in black hole formation, accompanied by emissions of both gravitational and electromagnetic radiation. A similar instability and two-stage pattern is observed in all charged rotating boson stars analyzed. However, all of these configurations become stable when self-interactions are introduced.
\end{abstract}
\date{\today}

\maketitle


\section{Introduction}
\label{Sec:intro}
We are in an unprecedented era of astrophysics, where an influx of new observations   
from gravitational waves and electromagnetic signals to neutrino detections~\cite{Janka:2012wk,IceCube:2013low,LIGOScientific:2017vwq,Kocevski:2017liw}, 
is reshaping our understanding of the Universe.
Gravitational wave astronomy began in 2015 with the groundbreaking detection of gravitational
waves from the merger of a binary black hole system, event GW150914~\cite{LIGOScientific:2016aoc}. Two years later,
the simultaneous detection of gravitational waves from a binary neutron star inspiral, event GW170817, along with
its post-merger electromagnetic counterpart, marked the dawn of multimessenger
astronomy~\cite{LIGOScientific:2017vwq,Cowperthwaite:2017dyu}.

Numerical simulations play an essential role in multimessenger astronomy, as they generate the waveforms needed to interpret 
signals detected by the LIGO/Virgo/KAGRA collaboration, and anticipate those from upcoming detectors like the Einstein Telescope 
and LISA~\cite{LISAConsortiumWaveformWorkingGroup:2023arg,Hild:2010id}. Besides, these simulations also predict electromagnetic
counterparts, such as gamma-ray bursts (GRBs) and electromagnetic luminosity~\cite{Paschalidis:2014qra,Ruiz:2016rai,Kiuchi:2023obe,Liebling:2016orx}, while modeling the fraction of ejected mass 
responsible for kilonova events~\cite{Just:2023wtj,Metzger:2014ila,Curtis:2021guz}, which can be observed through X-ray 
instruments like CHANDRA and XMM-Newton~(see~e.g.~\cite{Haggard:2017qne,DAvanzo:2018zyz}).

However, despite the advances made through multimessenger observations combined with numerical simulations, there are still open questions 
regarding the nature of the progenitors of some events, such as GW190814~\cite{LIGOScientific:2020zkf}.
This event involved the merger of a black hole with an object in the so-called mass gap~\cite{Barr:2024wwl,2020ApJ...897..100V}, of around $2.6M_\odot$, 
too massive to be a typical neutron star but too light to be classified as a black hole. Additionally, it has been shown
that GW190521 is consistent not only with the merger of a black hole binary but also with the collision of vector boson (Proca) stars~\cite{CalderonBustillo:2020fyi,CalderonBustillo:2022cja}, pointing to a degeneracy of the possible sources. These findings suggest the potential involvement of exotic neutron stars or other unknown compact objects. The detection of 
unusual black hole masses and possible intermediate-mass black holes further emphasizes significant gaps in our understanding of 
gravitational source in the Universe, emphasizing that gravitational wave astronomy continues to raise important, unresolved 
questions in  astrophysics and general relativity.

Exotic compact objects (ECOs), that include quark stars, and strange stars (stars containing strange quarks),
have also been proposed as alternative explanations for several ``unexplained'' electromagnetic phenomena, including fast radio bursts (FRBs) 
\cite{Dai:1998bb,Bombaci:2000cv,Zhang:2024xod}. For instance, magnetized exotic compact objects may undergo oscillations or instabilities that 
trigger violent energy releases, which could manifest as FRBs. When interacting or merging with other compact objects, the strong coupling 
between scalar and electromagnetic fields might cause bursts of radiation, explaining the millisecond-duration nature of FRBs. Moreover, the 
intense magnetic fields associated with gauged boson stars could lead to magnetospheric reconnections, similar to mechanisms proposed for 
neutron stars~\cite{Cheong:2024stz}, further supporting their role as potential FRB sources.
Beyond FRBs, exotic compact objects could also provide explanations for other astrophysical anomalies, such as
superluminous supernovae and GRBs. Some theories suggest that exotic stars could form in extreme astrophysical events,
such as neutron star mergers or collapses, releasing enormous amounts of electromagnetic energy that are detectable
across various wavelengths of the electromagnetic spectrum~\cite{Ruiz:2017inq,Ponce:2014sza}.

Furthermore, advances in theoretical modeling and computational techniques have provided new ways to explore the existence of alternative compact objects~(see e.g.~\cite{Bezares:2024btu} for a recently review). The study of self-gravitating objects composed of scalar fields has proven especially fruitful. The relative simplicity of the Einstein-Klein-Gordon system allows for the exploration of novel geometries, which can be used to explain several astrophysical observations. This approach not only challenges the traditional black hole paradigm but could also help to solve some of the puzzles mentioned above.
In the case of a complex scalar field, these self-gravitating and horizonless objects are known as boson stars (for reviews, see \cite{Shnir:2022lba,Liebling:2012fv}). When minimally coupled to gravity, the dynamical properties of these solutions are primarily determined by the choice of scalar field potential and whether angular momentum or non-spherical morphologies are introduced. The stability of these families of solutions has been a central focus of discussion for some time \cite{Gleiser:1988rq,Lee:1988av,Seidel:1990jh,Balakrishna:1997ej,Sanchis-Gual:2019ljs,Sanchis-Gual:2021edp,Siemonsen:2020hcg,Alcubierre:2021mvs,Jaramillo:2020rsv,Ildefonso:2023qty}. However, the stability of solutions when the scalar field interacts with a Maxwell field has been less thoroughly studied. The interplay between scalar and electromagnetic interactions in the $\mathrm{U}(1)$-gauged model of complex scalar fields with minimal electromagnetic coupling opens up the possibility of constructing explicit, self-consistent charged and magnetized spacetimes \cite{Jetzer:1989av,Delgado:2016jxq,Herdeiro:2020xmb,Jaramillo:2023pny,Jaramillo:2022gcq,Kunz:2023qfg}, leading to a rich array of phenomena \cite{Sanchis-Gual:2015lje,Zhang:2023qtn,Sanchis-Gual:2022zsr}.

In this work, we present a comprehensive analysis of  gauged scalar field self-gravitating solitons, commonly known as charged boson stars. We begin with the spherical case, which has previously been considered stable~\cite{Jetzer:1989av,Jetzer:1989us}. 
Through numerical evolution, we confirm that stability persists in a manner similar to that of free-field neutral boson stars \cite{Kaup:1968zz,Ruffini:1969qy,Gleiser:1988rq,Lee:1988av,Seidel:1990jh}, where an entire branch of solutions is present. 
We then examine the axisymmetric (spinning) case, where the motion of the system generates both electric and magnetic components. While uncharged objects were known to be unstable, 
our results show that the presence of an electric charge intensifies this instability and plays a role on  how the star collapses to a black hole.
Next, we introduce a specific configuration in which scalar fields, that interact only through gravity, are superimposed to form a static object with a toroidal morphology \cite{Jaramillo:2022gcq}. This object carries no net charge and, consequently, does not generate an electric field, but it does produce a poloidal magnetic field. This model serves as a simple analog to objects observed in nature, electrically neutral yet exhibiting a magnetic field. However, this configuration is also unstable.  Instabilities emerge more rapidly as the individual electric charge of each torus increases.

Motivated by the stability properties of the above configurations, we  probe the case where a self-interaction term is included in the scalar potential. Previous studies have shown that rotating boson stars with self-interaction are numerically stable~\cite{Siemonsen:2020hcg,DiGiovanni:2020ror}. As expected we find that  a rotating, uncharged scalar torus with both squared and quartic terms in the scalar potential remains stable over 
long evolution times. In particular, our configuration remains stable for $10^4$ times the characteristic oscillation time of the scalar field. 
Following this, we investigate the impact of the electromagnetic field on stability by gauging the scalar field for various charge 
values. We  then explore the non-spinning magnetic and electrically neutral star described above. This configuration
also proves stable across several values of the charge. In cases where the system becomes unstable and undergoes gravitational collapse 
into  a black hole, we observe emissions of gravitational, electromagnetic, and scalar radiation. Therefore, we a provide comprehensive 
analysis of scenarios involving both scalar and electromagnetic fields, offering insights into their stability and collapse dynamics.

The remainder of the paper is organized as follows. In Sec.~\ref{sec:Gau_sc_field_coupled}  we  briefly review  the theory 
of gauged scalar field coupled to gravity and discuss the families of stationary asymptotically flat configurations that can be constructed within this physical model. In addition we describe some
of the diagnostics we monitor to verify the reliability of our numerical simulations and outline our numerical methods. In Sec.~\ref{sec:single_field} we  probe the stability properties of
both spherical and axisymmetric charged boson star
configurations. In Sec.~\ref{sec:two_fields}, we address the stability of electrically neutral magnetized boson stars, while in
\ref{Sec:stable-conf}  we describe the evolution of self-interacting axisymmetric stars. Finally, we summarize our results and 
conclude in Sec.~\ref{sec:conclusions}. Throughout the paper we adopt units in which the speed of light $c$, gravitational constant $G$, 
and the vacuum permeability $\mu_0$ are equal to  one.

%
\section{Gauged scalar field coupled to gravity}
\label{sec:Gau_sc_field_coupled}
\subsection{Field Equations}
We begin with the action of a single charged scalar field $\Phi$ with potential $V(|\Phi|)$ and minimally coupled to gravity and standard electromagnetism,
\begin{equation}\label{eq:action}
S=\int d^4 x\,\sqrt{-g}\,\left[\frac{1}{16\pi }R-g^{\mu\nu}\mathcal{D}_\mu\Phi\mathcal{D}_\nu\Phi^*-V(|\Phi|)-\frac{1}{4}F_{\mu\nu}F^{\mu\nu}\right],
\end{equation}
where $F_{\mu\nu}=\partial_\mu A_\nu-\partial_\nu A_\mu$ is the Faraday tensor, $g_{\mu\nu}$ is the metric of the spacetime, $R$ 
is the Ricci scalar, and the covariant derivative operator is defined as $\mathcal{D}_\mu=\nabla_\mu+i\,q\,A_\mu$, where $q$ is 
the gauge coupling constant (also known as the scalar field charge parameter). Here  and 
throughout, we adopt the convention~$\mathcal{D}_\nu\Phi^*:=(\mathcal{D}_\nu\Phi)^*$.

Variations with respect to $g_{\mu\nu}$ leads to the Einstein field equations
\begin{equation}
\label{eq:einstein}
R_{\mu\nu}-\frac{1}{2}R g_{\mu\nu}=8\pi T_{\mu\nu} \,,
\end{equation}
where the stress energy tensors  is given by
\begin{equation}
T_{\mu\nu}=2\mathcal{D}_{(\mu}\Phi \, \mathcal{D}_{\nu)}\Phi^*-g_{\mu\nu}\left(\mathcal{D}^{\beta}\Phi \, \mathcal{D}_{\beta}\Phi^*+V(|\Phi|)\right)
+F_{\mu\sigma} F_{\nu\lambda}g^{\sigma\lambda}-\frac{1}{4} g_{\mu\nu} F_{\alpha\beta} F^{\alpha\beta}.
\label{Eq:tmunu}
\end{equation}

Variation with respect to $A_\mu$ leads to the Maxwell equations
sourced by the charged scalar fields which define a current four-vector $J^\mu$,
\begin{subequations}
\label{eq:maxwell}
  \begin{eqnarray}
    &&\nabla_\nu F^{\mu\nu}=J^\mu:=q J_\Phi^\mu\,,
    \label{eq:Maxwell0}\\
    &&J^\mu_\Phi:=i\left({\Phi}^*\mathcal{D}^\mu\Phi-\Phi \mathcal{D}^\mu{\Phi}^*\right),
  \end{eqnarray}
\end{subequations}
here $J_\Phi^\mu$ is a conserved current which arises from the gauge invariance of the scalar field which at the same time acts as the source $J^\mu$ of the electromagnetic field. Finally, variation with respect $\Phi$ leads to the Klein-Gordon equation
%
\begin{equation}\label{eq:kg}
\mathcal{D}^\mu \mathcal{D}_\mu \Phi=\frac{d V}{d|\Phi|^2}\Phi\,.
\end{equation}
%
\subsection{Stationary gauged boson stars}
\label{sec:stationary}
Gauged boson stars
are spatially localized solutions to the Einstein-Maxwell-Klein-Gordon system, described by Eqs.~\eqref{eq:einstein}, \eqref{eq:maxwell}, \eqref{eq:kg}. 
As described in~Sec.~\ref{Sec:intro} self-consistent solutions of this type, involving scalar fields coupled to a gauge field, have been numerically 
constructed and studied under assumptions of spherical symmetry, axial symmetry, and non-trivial electromagnetic fields.
We now present a more detailed description of the three gauged boson star models mentioned above.  In the following analysis, we will
focus on configurations consisting of free scalar fields, specifically a massive scalar potential without self-interaction
\begin{equation}
    V(|\Phi|) = \mu^2|\Phi|^2\,. 
    \label{eq:Vphi}
\end{equation}
We note that both static and rotating charged stars have been successfully modeled using more general (polynomial) potentials~\cite{Brihaye:2009dx,Collodel:2019ohy}. 
For simplicity, we will first focus on the massive case, with a more general model incorporating a quartic self-interaction to be introduced later.
Since we are focused on stationary axisymmetric boson star solutions, we adopt the following ansatz for the scalar field
\begin{equation}\label{eq:ansatz_phi}
\Phi = \phi(r,\theta)e^{i(\omega t - m\varphi)}\,,
\end{equation}
where $m$ is a integer parameter, known as azimuthal harmonic index~\cite{Herdeiro:2015gia}, and  $\omega$  is the angular velocity of
the scalar field.
For the spacetime, we use the corresponding line element
\begin{equation}\label{eq:ansatz_metric}
g_{\mu\nu}dx^\mu dx^\nu = - e^{2 F_0(r,\theta)} dt^2 + e^{2 F_1(r,\theta)}\left(dr^2 + r^2 d\theta^2\right) + e^{2F_2(r,\theta)}r^2\sin^2\theta\left(d\varphi - W(r,\theta) dt\right)^2 \,,
\end{equation}
where $F_0(r,\theta)$, $F_1(r,\theta)$, $F_2(r,\theta)$, and $W(r,\theta)$ are  four unknown metric functions. Notice that this
axisymmetric line element which describes a circular spacetime is written in quasi-isotropic coordinates~\cite{Gourgoulhon:2010ju}.
Similarly,
we use the following ansatz for the electromagnetic field four-potential
\begin{equation}\label{eq:ansatz_A}
A_\mu dx^\mu = V(r,\theta)\,dt + C(r,\theta)\,d\varphi\,.
\end{equation}

Given the above assumptions, we can build the following boson star configurations:
%
%
\paragraph{\bf Static charged boson stars:} Spherical charged boson stars are achieved when rotation is disabled. This case corresponds to the choice $m=0$ 
and $\phi=\phi(r)$ for the scalar field, $F_0=F_0(r)$, $F_1(r)=F_2(r)$ and $W=0$ for the gravitational field, and $C=0$ and $V=V(r)$ for the electromagnetic field. 
The equilibrium solutions in this scenario form sequences characterized by the coupling constant $q$, which has an upper bound of approximately
$q=\sqrt{4\pi}$. 
This upper bound arises from the limit set by Coulomb repulsion. For further details and discussion, 
we refer the reader to~\cite{Jetzer:1989av,Pugliese:2013gsa,Lopez:2023phk,Jaramillo:2023lgk,Kleihaus:2009kr,Kumar:2014kna}.

The linear stability of charged boson stars was analyzed by Jetzer~in~\cite{Jetzer:1989us},  who demonstrated that, similar to neutral mini-boson stars, 
an unstable branch of configurations emerges once the total mass exceeds its first maximum. 
Therefore, all configurations from the Newtonian dilute limit up to the critical point are potentially stable, whereas configurations with nodes are unstable.
Another preliminary stability analysis of these solutions was conducted by López and Alcubierre~in~\cite{Lopez:2023phk}, who performed short-term evolutions of 
several ground-state charged boson star configurations. They found that configurations with densities below the critical value remain stable, with stability 
inferred primarily from numerical discretization errors.
Additionally, they identified super-critical configurations, which they construct and then analyze. However, these configurations do not connect to the $M = 0$, 
$\omega / \mu = 1$ limit and exhibit only a minimum, lacking a maximum. Through their full general relativistic evolutions in spherical symmetry, they 
concluded that these super-critical configurations are unstable.

%
\paragraph{\bf Rotating charged boson stars:}
Axisymmetric charged boson stars, configurations with $m=1$ were first constructed for the self-interacting Q-ball-like potential in 
\cite{Brihaye:2009dx},  and later for the free-field case in~\cite{Delgado:2016jxq}. In the latter, sequences of rotating charged configurations 
were shown to be mathematically connected to hairy, charged black hole solutions. These configurations were revisited, analyzed, and the families of 
solutions where extended in~\cite{Herdeiro:2021jgc} for the free-field case, and in~\cite{Collodel:2019ohy} for scenarios with non-vanishing self-interactions.
These configurations are obtained using the full ansatz for $\Phi$, $g_{\mu\nu}$ and $A_\mu$ introduced earlier. 
The resulting system of equations forms a set of partial differential equations in $r$ and $\theta$, 
making it more challenging to solve compared to the electrostatic case.
Similar to the spherically $m=0$ counterparts, these configurations also form a sequence of solutions with increasing coupling constant  $q$.
The dynamical properties of the neutral case have been extensively studied. However, to the best of our knowledge, no stability assessments have been reported when electromagnetic 
coupling is included.
When $q=0$, the stability properties of mini-boson stars are known. For the $m=0$ case, both linear~\cite{Gleiser:1988rq,Lee:1988av} and nonlinear~\cite{Seidel:1990jh,Balakrishna:1997ej} 
stability analyses have firmly established that solutions connected between the Newtonian $M\to0$ mass limit and a critical point of maximum mass are stable. In contrast, due to 
the complexity involved, spinning boson stars have only recently been investigated in nonlinear studies~\cite{Sanchis-Gual:2019ljs, DiGiovanni:2020ror, Siemonsen:2020hcg}.

Fully non-linear numerical simulations reported in~\cite{Sanchis-Gual:2019ljs} found that rotating neutral boson stars are unstable.\footnote{Earlier studies on the orbital collisions of boson stars and the collapse of spinning scalar clouds suggested
that these configurations do not form stable configurations with non-zero angular momentum \cite{Bezares:2017mzk,DiGiovanni:2020ror}.}
Eventually, the system develops a non-axisymmetric instability. It is now understood~\cite{Siemonsen:2020hcg} that stationary solutions persist for longer 
times as one approaches the $\omega \to \mu$ (Newtonian) limit, but the instability is not suppressed. However, it was found that introducing a second scalar 
field in the fundamental spherically symmetric star~\cite{Sanchis-Gual:2019ljs}, or incorporating self-interactions across a range of potentials~\cite{Siemonsen:2020hcg,Sanchis-Gual:2019ljs}, 
can effectively suppress the instability.
As a final remark on rotating single-field solutions, similar to the spherical case, the sequence of rotating boson star solutions forms a spiral-like curve in mass vs.
frequency diagrams. 
This curve has a maximum mass point that separates the ``potentially stable'' branch from the unstable one. When the evolution or perturbations are constrained to axisymmetry, the 
instability in the ``potentially stable'' branch does not manifest, allowing for the study of intriguing gravitational wave phenomena, as demonstrated in~\cite{Siemonsen:2024snb}.

\paragraph{\bf Magnetostatic boson stars:} 
These stars are constructed differently from the previous two cases, as this model is based on the action in Eq~~\eqref{eq:action}, incorporating two massive and non-interacting scalar fields 
with opposite charges. The action is now defined as
\begin{equation}\label{eq:action_two}
S=\int d^4 x\,\sqrt{-g} \,\left(\frac{1}{16\pi }R-g^{\mu\nu}\mathcal{D}^{(+)}_\mu\Phi_+\mathcal{D}^{(+)}_\nu\Phi_+^*-
\mu^2|\Phi_+|^2-g^{\mu\nu}\mathcal{D}^{(-)}_\mu\Phi_-\mathcal{D}^{(-)}_\nu\Phi^*_--\mu^2|\Phi_-|^2-\frac{1}{4}F_{\mu\nu}F^{\mu\nu}\right)\,,
\end{equation}
where the covariant derivative operators are defined as $\mathcal{D}^{(\pm)}_\mu=\nabla_\mu\pm i\,q\,A_\mu$.
The magnetostatic configurations represent equilibrium states where both the total electric charge and total angular momentum are set to zero by construction. To achieve this, in the spacetime metric ansatz in Eq.~\eqref{eq:ansatz_metric}, $W$ is set to zero, while in the electromagnetic field ansatz we now set the electric potential $V$ to zero. The sources are modeled as a superposition of two counter-rotating tori (in terms of energy density). In this framework, the ansatz for the scalar fields is 
\begin{equation}\label{eq:ansatz_phi_two}
\Phi_\pm = \phi(r,\theta)e^{i(\omega t \mp m\varphi)} \, .
\end{equation}

For $q=0$, the solutions correspond to the toroidal boson star configurations discussed in \cite{Sanchis-Gual:2021edp}, which are part of a broader family of 
multifield, multifrequency scalar field star solutions.  The introduction of electromagnetic coupling, resulting in the emergence of magnetic fields, was explored 
in~\cite{Jaramillo:2022gcq}. The stability and formation mechanisms of neutral configurations were also studied in~\cite{Sanchis-Gual:2021edp}, where it was found that, 
in general, toroidal boson stars are unstable.

The sequences of solutions for the three types of stars will be presented in Sec. \ref{sec:single_field}  for the single-field (i.e. electrostatic and rotating charged) 
boson stars and in Sec.~\ref{sec:two_fields} for the two-field configurations, and in~\ref{Sec:stable-conf} we discuss the stable configurations {\it via} the introduction 
of the self-interaction term.  However, before discussing these solutions and their physical stability properties, which are the 
central focus of this paper, we will first formulate the Einstein-Maxwell-Klein-Gordon system in a manner that ensures numerically-stable-code evolutions
based on a  given initial data configuration.
Additionally, we will outline the key physical quantities useful for characterizing these solutions.

\subsection{3+1 decomposition}
We  decompose the metric using the $3+1$ decomposition,
\begin{equation}
ds^2=-\alpha^2 dt^2 + \gamma_{ij} (dx^i + \beta^i dt)(dx^j + \beta^j dt) \,,
\end{equation}
to formulate the Einstein-Maxwell-Klein-Gordon system as a Cauchy problem. In our numerical simulations we use the  Baumgarte-Shapiro-Shibata-Nakamura-Oohara-Kojima 
(BSSNOK)  formulation~\cite{Nakamura:1987zz,Shibata:1995we,Baumgarte:1998te} (see \cite{Corichi:1991qqo, alcubierre2008introduction} for an introduction to the formulation).
Here $\alpha$ and $\beta^i$ are the lapse function and the shift vector, respectively, and $\gamma_{ij}$ is the induced metric on the three-dimensional family of 
spacelike hypersurfaces~$\Sigma_t$ of constant coordinate $t$ in which the spacetime is foliated. However, for discussion purposes,
we begin by writing down the usual ADM system of equations. In the appendix \ref{sec:app:BSSN} the BSSNOK formulation couple to the 
electromagnetic field.
%
\subsubsection{Electromagnetic field}
For the evolution of the electromagnetic field we follow \cite{Zilhao:2015tya}, where a vectorial Proca field $X^\mu$, a massive generalization of the potential $A^\mu$, was considered.
It was useful to notice that the term $-\mu_V^2 X^\mu$ in \cite{Zilhao:2015tya} for the Proca equation can be identified with the $J^\mu$ source term in the Maxwell equations (and only there), and we continue with the 3+1 decomposition of the current.
Following the analogous approach in \cite{Zilhao:2015tya} for the 3+1 decomposition of the electromagnetic field and the standard split of the electromagnetic current \cite{Torres:2014fga}, we introduce
\begin{equation}\label{eq:DefAJ}
A_\mu=\A_\mu + n_\mu\A_\phi,    \qquad J_\mu=\mathcal{J}_\mu + n_\mu\rho_e\,,
\end{equation}
with $\A_k={\gamma^\mu}_k\,A_\mu$ and $\A_\phi=-n^\mu\,A_\mu$ the vector and scalar potentials and $\mathcal{J}_k={\gamma^\mu}_k J_\mu$ and $\rho_e=-n^\mu\,J_\mu$ the current density and charge density as measured by the Eulerian observers, $n^\mu$ is the unit normal vector to $\Sigma_t$. In the $3+1$ formulation of the electromagnetic field, it is also useful to define the electric and magnetic fields as measured by these observers, described by
\begin{equation}
\label{eq:DefEB}
E_{i} = \gamma^{\mu}{}_{i} F_{\mu\nu} n^{\nu}
\,,\quad
B_{i} = \gamma^{\mu}{}_{i} \,^{\ast} F_{\mu\nu} n^{\nu} = \epsilon^{ijk} D_{j} A_{k}\,,
\end{equation}
where $^{\ast} F_{\mu\nu}$ is the Hodge dual of $ F_{\mu\nu}$, and
$D_\mu$ the covariant derivative compatible with the three metric. Finally, following~\cite{Zilhao:2015tya, Hilditch:2013sba, Gundlach:2005eh,Palenzuela:2009hx}, 
an additional variable $Z$ and a parameter $\kappa$ are introduced to help damp the Gauss constraint during the numerical evolution.
With all this in place, we can now formulate the system governing the evolution of the gravitational and electromagnetic fields as 
\begin{subequations}\label{eq:adm_equations}
\begin{align}
\label{eq:dtgamma}
\p_{t} \gamma_{ij} & = - 2 \alpha K_{ij} + \Lie_{\beta} \gamma_{ij}
,\\
\label{eq:dtKij}
\p_{t} K_{ij}      & = - D_{i} D_{j} \alpha
        + \alpha \left( R_{ij} - 2 K_{ik} K^{k}{}_{j} + K K_{ij} \right)
        + \Lie_{\beta} K_{ij} 
\nonumber \\ &
        \quad +4\pi \alpha \left[ (S-\rho) \gamma_{ij} - 2 S_{ij} \right],\\
\label{eq:dtAi}
\p_{t} \A_{i}      & =  - \alpha \left( E_{i} + D_{i} \Aphi \right) - \Aphi D_{i}\alpha + \Lie_{\beta} \A_{i}
,\\
\label{eq:dtE}
\p_{t} E^{i}       & =
        \alpha \left( K E^{i} + D^{i} Z - \mathcal{J}^{i}
                + \epsilon^{ijk} D_{j} B_{k} \right)
        - \epsilon^{ijk} B_{j} D_{k}\alpha
        + \Lie_{\beta} E^{i},\\
\label{eq:dtAphi}
\p_{t} \Aphi  & =  - \A^{i} D_{i} \alpha
        + \alpha \left( K \Aphi - D_{i} \A^{i} - Z \right)
        + \Lie_{\beta} \Aphi ,\\
\label{eq:dtZ}
\p_{t} Z          & =  \alpha \left( D_{i} E^{i} -  \rho_e - \kappa Z \right)
        + \Lie_{\beta} Z\,,
\end{align}
\end{subequations}
where $K_{ij}$ is the extrinsic curvature, $\Lie_{\beta}$ is the Lie derivative along $\beta^i$, and the gravitational source terms 
are given by the 3+1 projections of the total energy-momentum tensor in Eq.~\eqref{Eq:tmunu},
\begin{equation}
  \label{eq:source}
   \begin{aligned}
  \rho = n^\alpha n^\beta T_{\alpha\beta}\, , \quad
  P_i = - \gamma_i{}^\alpha n^\beta T_{\alpha \beta}\, , \quad
  S_{ij} = \gamma^\alpha{}_i \gamma^\beta{}_j T_{\alpha \beta}\, , \quad
  S = \gamma^{ij} S_{ij}\, .
   \end{aligned}
\end{equation}
Note that the Lorenz condition $\nabla_\mu A^\mu=0$  was chosen in deriving Eqs.~\eqref{eq:dtAphi}-\eqref{eq:dtZ}.\footnote{For the Proca case the Lorenz condition is not a gauge choice but a requirement obtained from the equations of motion. However, in our case we have decided to keep this condition because it simplifies both the evolution of 
the scalar potential and the Klein-Gordon equation, as will be demonstrated below.}

%
\subsubsection{Charged scalar field}
Following~\cite{Torres:2014fga},  which analyzed the collapse of a charged scalar field, we begin by noting that the Klein-Gordon equation 
(Eq.~\eqref{eq:kg}) simplifies to $\Box \Phi + q \left( 2 i A^\mu \nabla_\mu \Phi - q A_\mu A^\mu \Phi \right)= \mu^2 \Phi$ in the Lorenz gauge. 
After this, it is possible to define a canonical momentum variable associated with the scalar field, analogous to the definition of the extrinsic 
curvature as described in~ \cite{Cunha:2017wao}),
\begin{equation}\label{eq:Kphi}
    K_{\Phi} = -\frac{1}{2\alpha}  \left( \partial_{t} - \Lie_{\beta} \right) \Phi \,.
\end{equation} 
In terms of this variable, the Klein-Gordon system can be reduced to first order in time as
\begin{subequations}
\begin{align}
  \p_{t} \Phi & = - 2 \alpha K_\Phi + \Lie_{\beta} \Phi\
                \,, \label{eq:dtPhi} \\
  \p_{t} K_\Phi &  = \alpha \left( K K_{\Phi} - \frac{1}{2} \gamma^{ij} D_i \partial_j \Phi
                  + \frac{1}{2} \mu^2 \Phi \right)
                 - \frac{1}{2} \gamma^{ij} \partial_i \alpha \partial_j \Phi
                       + \Lie_{\beta} K_\Phi \nonumber \\
                     & \quad + q^2\frac{\alpha}{2}\left(\gamma^{ij}\A_i\A_j-\Aphi^2\right)\Phi - iq\alpha\left(\gamma^{ij}\A_j\partial_i\Phi-2\Aphi K_\Phi\right)\,.
                     \label{eq:dtKphi}
\end{align}
\end{subequations}
Therefore, the complete Einstein-Maxwell-Klein-Gordon system can be numerically evolved using Eqs.~\eqref{eq:dtgamma}-\eqref{eq:dtZ} and Eqs.~\eqref{eq:dtPhi}-\eqref{eq:dtKphi}.
Notice that this system is  subject to the Hamiltonian $\mathcal{H}$ and the momentum $\mathcal{M}_{i}$ constraints,
\begin{align}
\label{eq:Hamiltonian}
\mathcal{H} & \equiv R - K_{ij} K^{ij} + K^2 - 16 \pi \rho
       = 0\,,\\
\label{eq:momentumConstraint}
\mathcal{M}_{i} & \equiv D^{j} K_{ij} - D_{i} K 
        - 8\pi P_i
       = 0 \,.
\end{align}
Before concluding the initial value problem related to the decomposition of the Einstein-Maxwell-Klein-Gordon system, 
it is useful to express the sources of the Maxwell equations in terms of the canonical momentum variable defined in Eq.~\eqref{eq:Kphi} as
\begin{subequations}
\begin{align}
   \rho_e & = iq\left[\Phi^*\left(2K_\Phi+iq\Aphi\Phi\right) - \Phi\left(2K_\Phi^*-iq\Aphi\Phi^*\right) \right] \,, \\
  \mathcal{J}^i &  = iq\left[\Phi^*\left(\p^i\Phi+iq\A^i\Phi\right) - \Phi \left(\p^i\Phi^*-iq\A^i\Phi^*\right) \right]   \,. 
\end{align}
\end{subequations}
As mentioned above, we present the above evolution equations expressed in terms of the BSSNOK variables in Appendix~\ref{sec:app:BSSN}.

\subsection{Physical quantities}

To verify the reliability of our numerical simulations we monitor a number of several local and global diagnostics.
Considering also stationarity and axisymmetry, and asymptotically flatness, as is the case of the isolated unperturbed boson stars,
there are two global conserved charges, the mass and the angular momentum.
In this context the Komar mass can be obtained through~\cite{alcubierre2008introduction}
\begin{equation}\label{eq:komarM}
M = 2 \int_{\Sigma_t} \left( T_{\mu\nu} n^\mu \xi^\nu - \frac{1}{2} T n_\mu\xi^\mu\right)\,\sqrt{\gamma} d^3 x \,,
\end{equation}
where $\xi=\partial_t$ is the time-like 
Killing vector, and $T$ is the trace of the energy-momentum tensor.
Associated with the space-like Killing vector $\chi = \partial_\varphi$, the other global charge due to the symmetries of the spacetime is obtained in the form of angular momentum
\begin{equation}\label{eq:komarJ}
    J = -\int_{\Sigma_t} T_{\mu\nu}n^\mu\chi^\nu\sqrt{\gamma} d^3 x \, .
\end{equation}

When constructing the stationary solutions we employ also the asymptotic behaviour of the geometry to get the mass and angular momentum in order to check the accuracy of the code. 
For example, we verify that extracting this quantities from the asymptotic behaviour of the metric functions $g_{tt} = -1+2M/r + \mathcal{O}(1/r^2)$ and $g_{t\varphi} = 2J\sin^2\theta/r + \mathcal{O}(1/r^2)$ give consistent results. Also we transform Eq.~\eqref{eq:komarJ} to a volume integral and use the Einstein equations to test the accuracy in $J$ by comparing its value numerically calculated in both ways as in \cite{Grandclement:2014msa}. As an error indicator for $M$ we use the ADM mass as defined in Eq.~(16) of \cite{Jaramillo:2022gcq} and compare with the Komar mass, which is guaranteed to coincide in the stationary case.

Associated with the local $\mathrm{U}(1)$ symmetry of the gauged scalar field we have the Noether charge (see Eq.~\eqref{eq:Maxwell0})
\begin{equation}\label{eq:Q}
  Q = -\int n_\mu J_\Phi^\mu\sqrt{\gamma}d^3 x \, .
\end{equation}
By substituting the Noether current $J_\Phi$ by the electric current $J$  in the above expression,  we obtain the electric charge of the configuration.
From Eqs.~\eqref{eq:maxwell},  it follows that $Q_E = q\, Q$. As shown in~\cite{Collodel:2019ohy}, under the conditions for stationarity, the quantities $Q$ 
and $J$ are related.

The total electric charge can also be extracted from the asymptotic behaviour of the electromagnetic potential 
\begin{equation}
A_t = - \frac{Q_E}{4\pi\,r} + \mathcal{O}\left(\frac{1}{r^2}\right) \,.
\end{equation} 
We that in in our units the vacuum permittivity is one.
The magnetic dipole moment $\mu_m$ can be extracted from the other non-zero component of the 4-potential as 
\begin{equation}\label{eq:assymptoticAphi}
    A_\varphi = \frac{\mu_m\,\sin^2\theta}{4\pi\,r} +\mathcal{O}\left(\frac{1}{r^2}\right) \, .
\end{equation}

As the solutions we will consider next involve configurations with both the angular momentum and the charge, we have calculated their 
gyromagnetic ratio to quantify the differences between the horizonless objects presented and the well-known Kerr-Newman exact solution.
It is a remarkable fact that the electric charge, magnetic dipole moment and angular momentum of the Kerr-Newmann black holes are 
interrelated in such a way that the gyromagnetic ratio $\mathcal{G} = 2\mu_mM/(Q_E J)$ is equal to 2. Besides, we note that for the 
Kerr-Newmann  black hole, the quantity 
\begin{equation}
\label{eq:assymptoticDelta}
\Delta = \frac{M^2}{{Q_E^2}/{4\,\pi} + J^2/M^2}\,,
\end{equation}
is always greater than one, ensuring that an event horizon hides the curvature singularity \cite{Delgado:2016jxq}. 

%
\begin{figure}
    \includegraphics[width=0.5\textwidth]{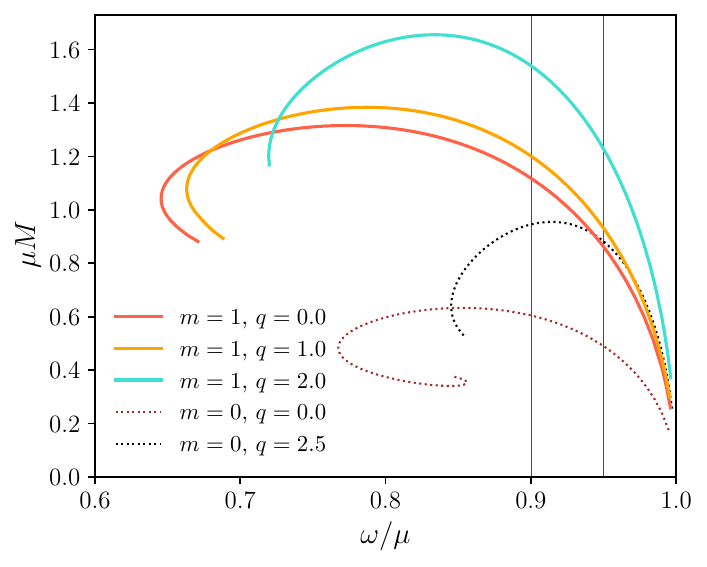}\includegraphics[width=0.5\textwidth]{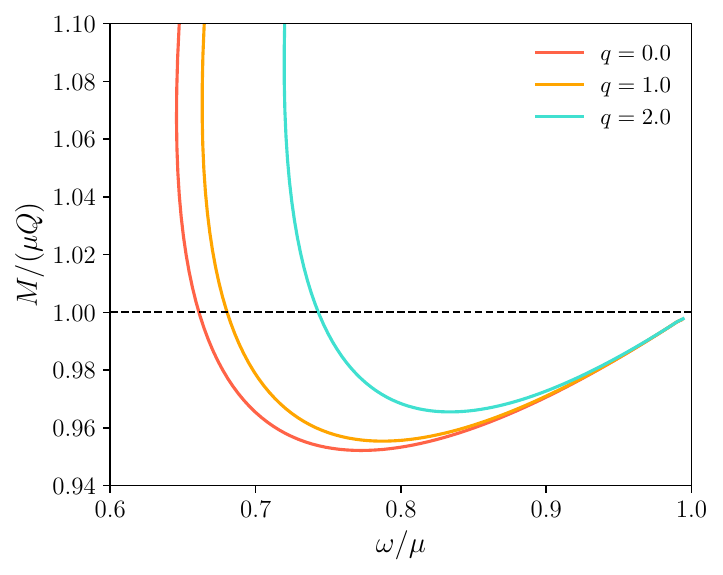}
    \caption{Sequences of charged and rotating boson stars. The left panel shows the total mass \textit{vs.} the scalar field frequency $\omega$ for 
    spherical ($m=0$) boson stars (dotted lines) and the axisymmetric, rotating ($m=1$) stars (continuous lines) for different values of 
    the gauge coupling constant $q$. 
    The right panel shows the mass-to-the-Noether-charge ratio for the $m=1$ configurations and 
    three different values of the constant $q$. Configurations with $M/Q$ above unity are expected to fragment and are therefore 
    potentially unstable. In all cases explored in this work, the change in sign of the binding energy occurs in the second branch (to the left of the critical mass configuration in these plots). The vertical lines in the left panel indicate the two frequencies considered in the dynamical analysis, see Table \ref{tab:electric_table}.}
    \label{fig:BSsequences}
\end{figure}
%

%
\subsection{Numerical approach}
\label{sec:Numerical methods}
\paragraph{\bf Initial data:} We solve the coupled Einstein-Maxwell-Klein-Gordon system for stationary and axisymmetric gauged boson stars, 
assuming the ansatz discussed in Sec.~\ref{sec:stationary}. We impose regularity throughout the spacetime, asymptotic flatness and use
\texttt{Kadath} library~\cite{Kadath,Grandclement:2009ju} coupled to the boson star initial data solver described in~\cite{Grandclement:2014msa}.
It employs the \texttt{Polar} space module, which guarantees the regularity of the fields at the coordinate center and allows for the compactification 
of the radial domain. For all configurations in this work, we employed $8$ radial domains with $17$ spectral coefficients in both  $r$ and $\theta$ 
coordinate. The Newton iteration is halted when the residual drops below $10^{-8}$. The specific equations used in 
the solver are detailed in Appendix~\ref{app:initial_data_spherical}, where we present the compact forms derived from applying the ansatz to the 
Einstein-Maxwell-Klein-Gordon system. Left panel in~Fig.~\ref{fig:BSsequences} displays the mass \textit{vs.} scalar field frequency $\omega$
for spherical ($m=0$) boson stars (dotted lines) and axisymmetric, rotating ($m=1$) stars (continuous lines) at different values of the gauge coupling 
constant $q$. According to classical physics arguments, the critical value of $q$ beyond which no further spherical (Newtonian) equilibrium configurations can  be found is around 
$q_{\rm max} = 3.545$ \cite{Jetzer:1989av, Pugliese:2013gsa}. 
Between the neutral case and $q=3.5$ (near the critical charge), the behavior of the solution families is well-defined. 
The maximum mass increases monotonically, and the frequency of the critical solution approaches the limit $\omega/\mu = 1$.
Right panel in Fig.~\ref{fig:BSsequences} shows the mass-to-Noether-charge ratio for the $m=1$ configurations for value of $q$ ranging between $q=0$ and $2.5$.
Configurations with $M/Q$  above unity are expected to fragment and are thus potentially unstable. In all cases explored in this work, the change 
in sign of the binding energy occurs in configurations beyond the critical mass solution.

\paragraph{\bf Evolution:} We evolve the gauged stars using the publicly available {\tt Einstein Toolkit} code~\cite{cactusweb}. In particular, we use 
the {\tt McLachlan} thorn~\cite{Brown:2008sb} to evolve the metric variables, which use the BSSNOK formulation coupled  with the moving-puncture gauge condition,
i.e. the ``Gamma-driver'' condition for the shift~$\beta^i$\cite{Alcubierre:2002kk}  and the ``1+log'' condition for the lapse $\alpha$~\cite{Bona:1994dr}. 
To evolve the gauged boson stars, we employ the newly developed 
{\tt Magnetoscalar} thorn, which is built upon the~{\tt Proca} and {\tt Scalar} thorns from the~{\tt Canuda} library~\cite{canuda_2023_7791842}. 
These thorns have been previously employed and described in~\cite{Zilhao:2015tya, Cunha:2017wao, Sanchis-Gual:2019ljs, Sanchis-Gual:2022zsr} 
for evolving charged scalar and electromagnetic fields. Additional details on the quantities needed for the evolution thorns are presented in 
Appendix~\ref{app:initial_data_rotating}.

 In all our simulations, we use a set of six nested refinement boxes, which is provided by the \texttt{Carpet} thorn \cite{Schnetter:2003rb}, centered 
at the star  and differing in size and resolution by factors of two. The finest level has a grid spacing of $\Delta x = 0.25$, and resolve the star 
radius by $\sim 70$ grid points.  We impose reflection equatorial symmetry. We ran some of the configurations in Table~\ref{tab:electric_table} 
at higher resolutions, increasing the resolution by a factor of $1.5$ and $2$, to ensure that our results are independent of resolution.
The time  integration is performed with the method of lines using a  
fourth-order accurate Runge-Kutta method, with a fixed time step and a Courant-Friedrichs-Lewy (CFL) factor of $0.225$. To mitigate high-frequency 
modes that may arise during the evolution of these gauged (charged/magnetized) boson stars, we apply a fifth-order Kreiss-Oliger dissipation with variable strength,  depending on the local Courant factor, defined as  $\epsilon_i=\rm{\epsilon}/ (2^{({\rm{max-levels}}-i - 1)})^5$, where $\epsilon$ 
is a constant,  $\epsilon_i$ represents the dissipation strength at the $i$-th refinement level, and ${\rm{max-levels}}$ is the total number of 
refinement levels employed during the evolution,  which is six in our case~(for details~see~\cite{Bozzola:2021elc}). We empirically found that 
setting $\epsilon=0.05$ ensures the  stability in all our  configurations. We employ radiative boundary conditions for all the evolve variables. 

%
\section{Spherical and rotating charged stars} 
\label{sec:single_field}

We begin by presenting our results for the case of single-field gravitational solitons within the framework of the theory described in~Eq.~(\ref{eq:action}), 
focusing first on the spherical ($m=0$) case, which serves as a validation of our numerical implementation of the {\tt Magnetoscalar} thorn. 
The first rows of Table~\ref{tab:electric_table} display several global quantities for configurations within the spherically 
symmetric families of solutions (see Fig.~\ref{fig:BSsequences}). 

As previously discussed, linear theory predicts that all spherical solutions, except for those lying between the $\phi \to 0$  ($\omega = \mu$, $M = 0$) and the point
of maximum mass $M$,  are expected to be unstable.
To probe the  dynamical stability properties of some of these configurations (see Table~\ref{tab:electric_table})  we evolve them in a full non-linear regimen 
up to $t\sim 10^4/\mu$, allowing the initial  data to be  perturbed only by truncation errors.
\begin{table}
  \centering
  \begin{tabular}{l|cccccc|c}
    \hline\hline
    $m$ & $\omega/\mu$ & $q$ & $\mu M$& $\mu^2Q$& $\max\phi$ & $\mu^2 \mu_m$ & Remnant  \\ \hline
    $0$ & $0.95$       & 0.0 & 0.490  & 0.497   & 0.0218     & 0.0           & BS       \\
    $0$ & $0.95$       & 0.5 & 0.499  & 0.507   & 0.0221     & 0.0           & BS       \\
    $0$ & $0.95$       & 1.0 & 0.528  & 0.536   & 0.0229     & 0.0           & BS       \\
    $0$ & $0.95$       & 1.5 & 0.584  & 0.593   & 0.0244     & 0.0           & BS       \\
    $0$ & $0.95$       & 2.0 & 0.686  & 0.696   & 0.0272     & 0.0           & BS       \\
    $0$ & $0.95$       & 2.5 & 0.882  & 0.894   & 0.0328     & 0.0           & BS       \\\hline
    $1$ & $0.9$        & 0.0 & 1.119  & 1.152   & 0.0226     & 0.0           & BH       \\
    $1$ & $0.9$        & 0.5 & 1.138  & 1.173   & 0.0229     & 0.375         & BH       \\
    $1$ & $0.9$        & 1.0 & 1.201  & 1.237   & 0.0237     & 0.802         & BH       \\
    $1$ & $0.9$        & 1.5 & 1.323  & 1.362   & 0.0254     & 1.356         & BH       \\
    $1$ & $0.9$        & 2.0 & 1.540  & 1.583   & 0.0284     & 2.191         & BH       \\
    $1$ & $0.9$        & 2.5 & 1.934  & 1.983   & 0.0345     & 3.701         & BH       \\
    $1$ & $0.95$       & 0.0 & 0.862  & 0.876   & 0.0109     & 0.0           & Mi        \\
    $1$ & $0.95$       & 0.5 & 0.879  & 0.893   & 0.0109     & 0.255         & Mi        \\
    $1$ & $0.95$       & 1.0 & 0.932  & 0.947   & 0.0114     & 0.545         & Mi        \\
    $1$ & $0.95$       & 1.5 & 1.036  & 1.053   & 0.0121     & 0.922         & Mi        \\
    $1$ & $0.95$       & 2.0 & 1.228  & 1.248   & 0.0134     & 1.496         & Mi        \\
    $1$ & $0.95$       & 2.5 & 1.611  & 1.635   & 0.0159     & 2.581         & BH       \\
    $1$ & $0.95$       & 3.0 & 2.567  & 2.601   & 0.0224     & 5.584         & BH       \\
    $1$ & $0.95$       & 3.2 & 3.408  & 3.444   & 0.0299     & 8.798         & BH\\
    \hline\hline
  \end{tabular}
  \caption{Selected configurations of single-field boson stars, spherical ($m=0$) or rotating ($m=1$) for two values of the scalar field frequency. 
  We present the final configuration, labeled {\it Remnant}, being the initial boson star (BS) (neutral, $q=0$ or charged, $q \neq 0$), black hole (BH) ({\it idem}) or migration to a not yet relaxed horizonless localized solution (Mi), for several combinations of the configuration parameters, namely the scalar field normalized frequency, $\omega/\mu$, the charge, $q$, the normalized total mass of the star, $\mu\,M$, the normalized total charge, $\mu^2\,Q$ , the absolute maximum of the scalar field ${\rm max} \phi$ and the normalized magnetic moment, $\mu^2\,\mu_m$.}
  \label{tab:electric_table}
\end{table}
As anticipated from Fig.~\ref{fig:BSsequences}, increasing the value of the coupling constant $q$ shifts the maximum mass to higher values of
$\omega$.  All stable configurations  lie within the first branch.
When exploring random ground state configurations in the second
branch (between the maximum mass and the configuration with the global minimum of $\omega$), we observe that the perturbation induced by truncation error is enough to seed the expected 
instability, confirming the linear results in~\cite{Jetzer:1989us}.  

\begin{figure}
    \includegraphics[width=0.8\textwidth]{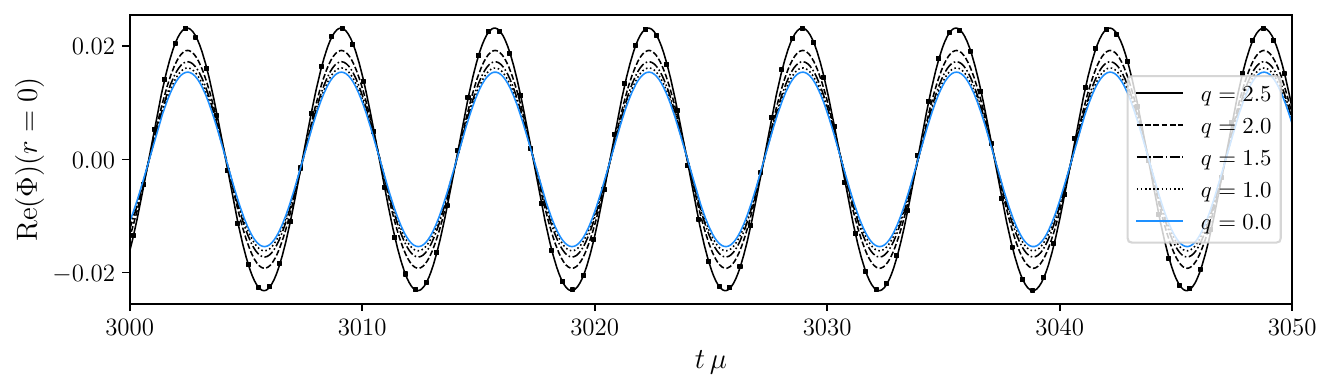}\\
    \hspace{0.2cm}\includegraphics[width=0.79\textwidth]{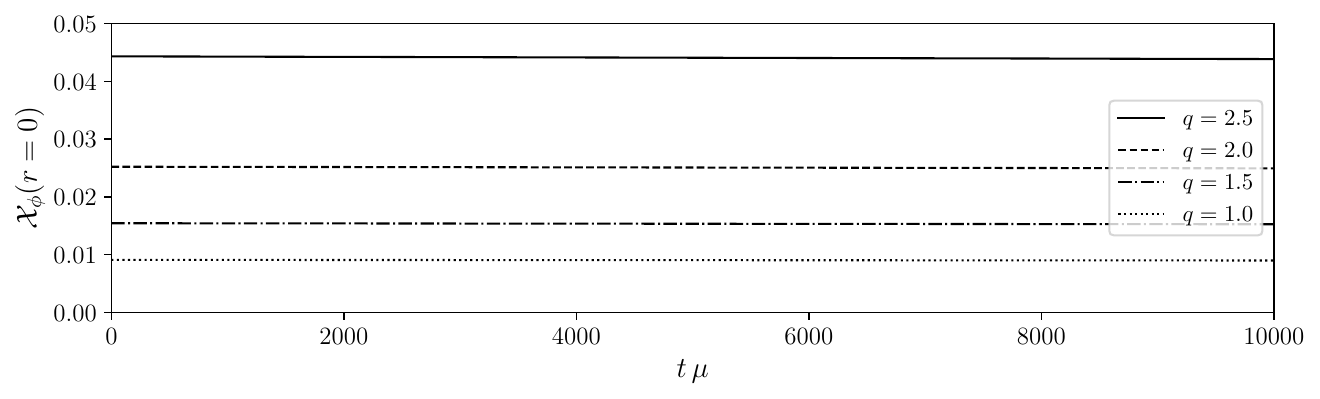}
    \caption{Evolution of the charged spherical star with $\omega=0.95\mu$ and several values of the coupling constant $q$. 
    Top panel highlights the periodicity profile of the real part of the scalar field $\re(\Phi)$ at center of the star along $\Delta t\,\mu=50$, though its behavior is the same during the whole 
    evolution. For comparison,  the top panel shows the functions $\max(|\Phi|)\cos(\omega t)$ (dotted curve) while in the case $q=2.5$ we explicitly show the numerical data (Bottom panel) 
    Electric potential. Showing the stability of the spherical boson star. 
    }
    \label{fig:RePhi}
\end{figure}

\begin{figure}
    \includegraphics[width=0.79\textwidth]{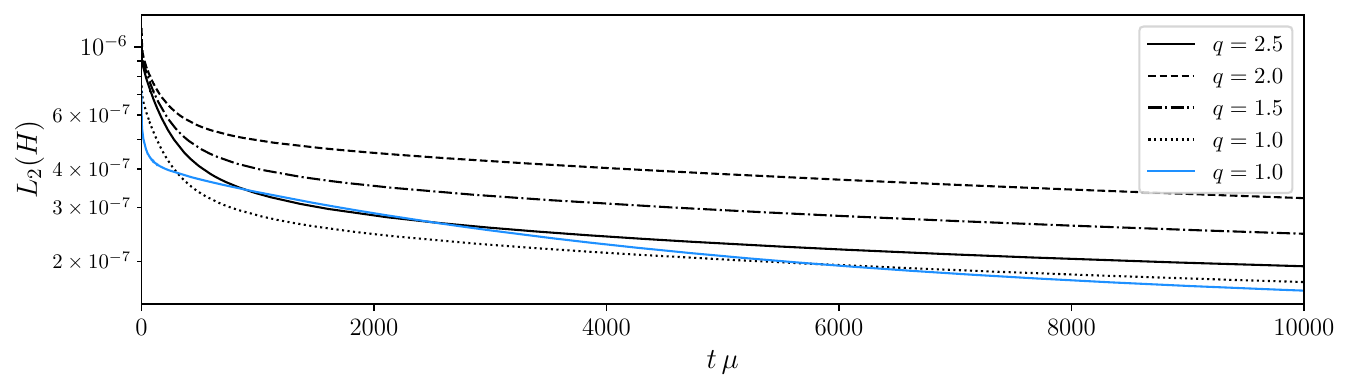}\\
    \hspace{0.2cm}\includegraphics[width=0.79\textwidth]{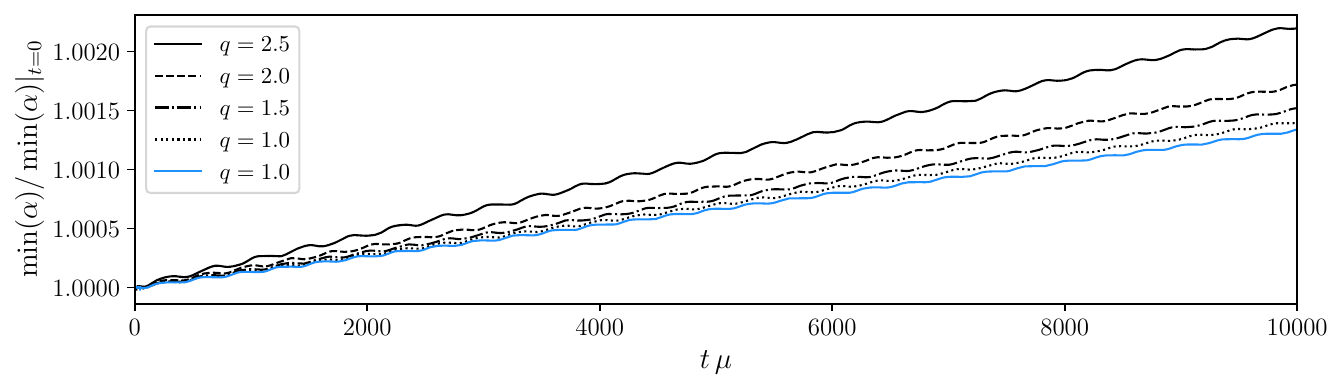}
    \caption{Charged spherical star with $\omega=0.95\mu$ and five different values of $q$. Top panel displays  the evolution of the $L_2$ norm of the Hamiltonian constraint, 
    while  the bottom panel shows the evolution of the absolute minimum of the lapse function. As expected the stable spherical stars oscillates during the whole evolution~\cite{Jetzer:1989us}.
    }
    \label{fig:Geometry_spherical}
\end{figure}

Fig.~\ref{fig:RePhi} displays the real part of the scalar field and the electric potential evaluated at the center of the star 
for charged stable spherical configuration with  $\omega=0.95\mu$. We find excellent agreement with the stationary solution $\Phi(0,t) 
= \phi(0)e^{i\omega t}$ and the same conclusion can be obtained analyzing other parts of the star. For comparison purposes,  the top panel in Fig.~\ref{fig:RePhi} shows the numerical values of the central value of~$\re(\Phi)$ for the 
configuration  with $q=2.5$, compared to the function $\max(|\Phi|)\cos(\omega t)$. This fitting procedure shows similarly good agreement across the other $q$ cases.
The reliability of the previous results is further supported from a gravitational perspective, as the Hamiltonian constraint, showed in Fig.~\ref{fig:Geometry_spherical}, 
remains well-bounded throughout the evolution. The bottom panel Fig.~\ref{fig:Geometry_spherical} shows the evolution of the lapse normalized but its initial value. We 
observe that the minimum value of the lapse slight increases by less than $ 0.5\%$ what may indicate a small change in the mass of the bulk of the star. This behavior 
has been observed in the evolution of other stable compact objects remnants such as neutron stars~\cite{Ruiz:2021gsv} and it is likely due to  numerical dissipation.

Having tested our numerical implementation through the  the spherical solutions, and confirmed the reliability of the evolutions. 
We now shift our focus to the spinning (axisymmetric) $m=1$ solutions, 
which are known to
be  unstable in the neutral charge limit.\footnote{The $m \geq 2$ cases are also known to be unstable~\cite{Siemonsen:2020hcg}, and no values of self-interaction 
parameters are currently known to stabilize the $m=2$ case.} 

We begin by describing the properties of the stationary solutions obtained from solving the elliptic system detailed in Appendix~\ref{app:initial_data_rotating}. 
Table~\ref{tab:electric_table} presents specific solutions for two selected frequencies, $\omega = 0.9\mu$ and $\omega = 0.95\mu$. For each frequency, we explore 
charge values ranging from $q = 0$ to $q = 2.5$, with the range extending up to $q = 3.2$ for $\omega = 0.95\mu$. 
As previously observed in~\cite{Delgado:2016jxq}, the mass of the configurations increases with the  electromagnetic coupling constant $q$, while the range of possible 
frequencies $\omega$ narrows as $q$ grows. 
This trend is depicted in the left panel of Fig.~\ref{fig:BSsequences}, where the mass of a sequence of solutions is shown for both rotating ($m=1$) and spherical ($m=0$)~cases. 
In the right panel the  mass-to-the-Noether-charge ratio $M/Q$~is plotted, serving as an indicator of the binding energy, with $M/(\mu Q) = 1 + E_B/(\mu Q)$.
The system is gravitationally bound when $M/(\mu Q)<1$. From the right panel of Fig.~\ref{fig:BSsequences}, we infer that all configurations within the first 
branch of solutions are gravitationally bound, making them unlikely to fragment under small perturbations.

Rotating solutions ($m = 1$) exhibit angular momentum. As shown in \cite{Delgado:2016jxq} and \cite{Collodel:2019ohy}, 
the angular momentum and the Noether/electromagnetic charge are related  as
 \begin{equation}
    J=m\,Q=m\, \frac{Q_E}{q}\, .
\end{equation}
These identities, which can be derived using the stationary ansatz described in Sec.~\ref{sec:stationary}, provide an additional accuracy check for the code. We employed this error indicator (together with the Komar/ADM test) to verify the self-consistency of the solutions generated by the solver.

\begin{figure}
    \includegraphics[width=0.5\textwidth]{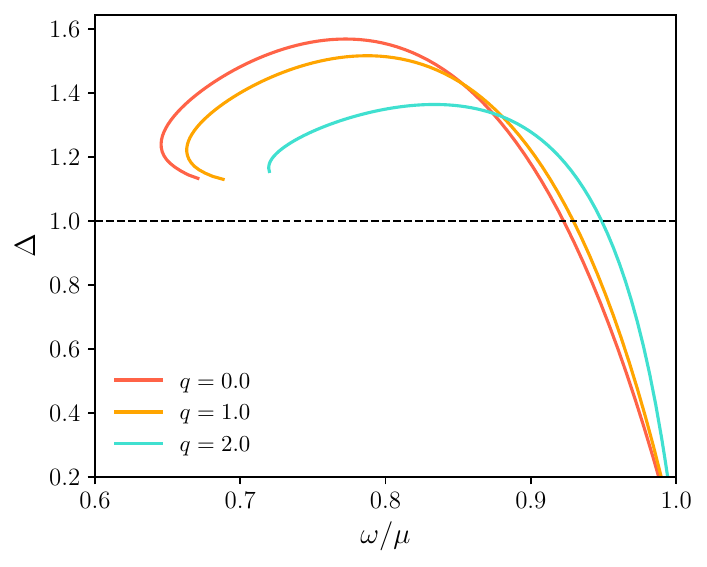}\includegraphics[width=0.485\textwidth]{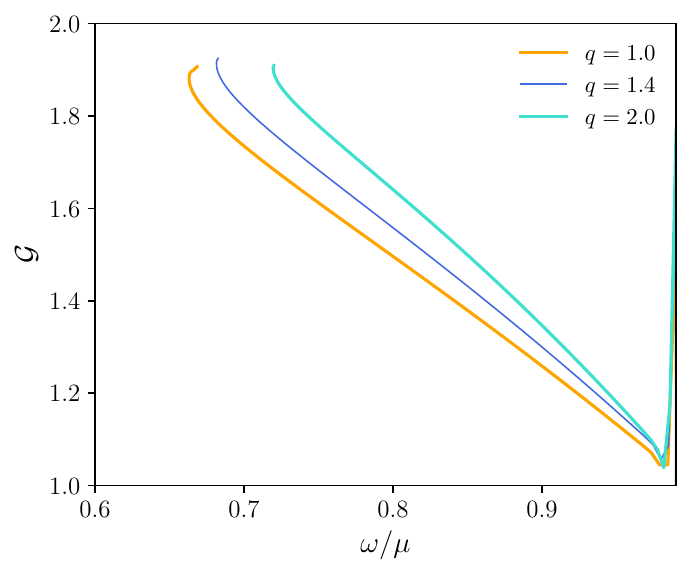}
    \caption{Charged rotating ($m=1$) stars with different values of the coupling constant~$q$. Left panel displays  the
    $\Delta$ function (defined in Eq.~\eqref{eq:assymptoticDelta}), which satisfies $\Delta\geq 1$ for Kerr-Newman electro-vacuum black holes.
    However, for the obtained boson star configurations we observe cases with $\Delta<1$, in particular, in regions close to the $\omega=\mu$. 
    Right panel shows the gyromagnetic radio for three charged configurations. From the magnetic components of the electromagnetic four-potential 
    we extract the magnetic moment according to Eq.~\eqref{eq:assymptoticAphi} and with this information we calculate the gyromagnetic ratio. 
    The gyromagnetic ratio of the Kerr-Newman family is $\Delta=2$.}
    \label{fig:Delta}
\end{figure}

We calculate the  gyromagnetic ratio  $\mathcal{G}$ and the extremality Kerr-Newman bound $\Delta$~(see~Eq.~(\ref{eq:assymptoticDelta})).
These quantities show significant differences compared to the electro-vacuum black hole solutions. As seen in the left panel of Fig.~\ref{fig:Delta}, the charged, rotating 
boson star satisfies the sub-extremal Kerr-Newman condition $\Delta > 1$ in some regions of the parameter space. Besides, the gyromagnetic ratio $\mathcal{G}$ is consistently 
lower than the Kerr-Newman value $\mathcal{G}=2$.

Having identified the families of rotating configurations, we evolve the configurations  listed in the second part of 
Table~\ref{tab:electric_table}. Vertical lines in the left panel of Fig.~\ref{fig:BSsequences} marks these frequencies 
($\omega = 0.9\mu$ and $\omega = 0.95\mu$).  Interestingly, although all these cases correspond to configurations 
within the first branch, which are expected to be stable, we find that all of them are unstable, in contrast to the quartic self-interaction 
potential (see below), which induces a positive (repulsive) force and has been shown to stabilize solutions within the first branch \cite{Siemonsen:2020hcg}.
The backreaction of the charged field, which generates an electric field in these configurations, also has a repulsive effect. However, the stability properties of charged 
boson stars, where gravitational, electromagnetic, and scalar fields are coupled, extend beyond simple analogies of attractive or repulsive forces. 
Since all the solutions are constructed as equilibrium states, a comprehensive stability analysis, whether linear or nonlinear, is necessary to draw definitive conclusions.

\begin{figure}
\subfigure[~~$q=0$]{
\includegraphics[height=0.135\textheight]{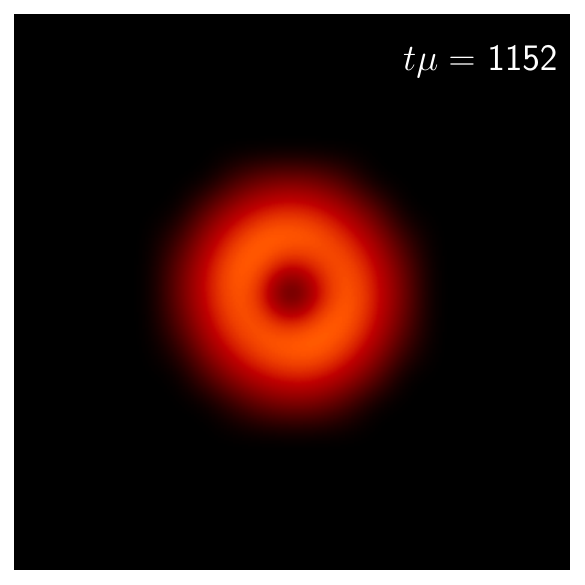}\hspace{-0.1cm}
    \includegraphics[height=0.135\textheight]{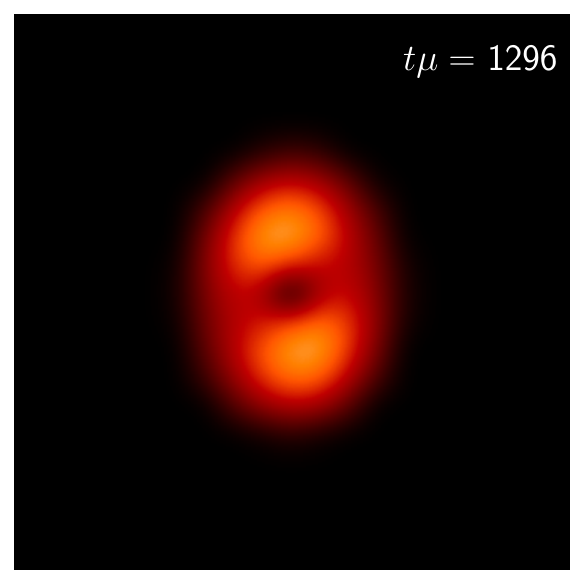}\hspace{-0.1cm}
    \includegraphics[height=0.135\textheight]{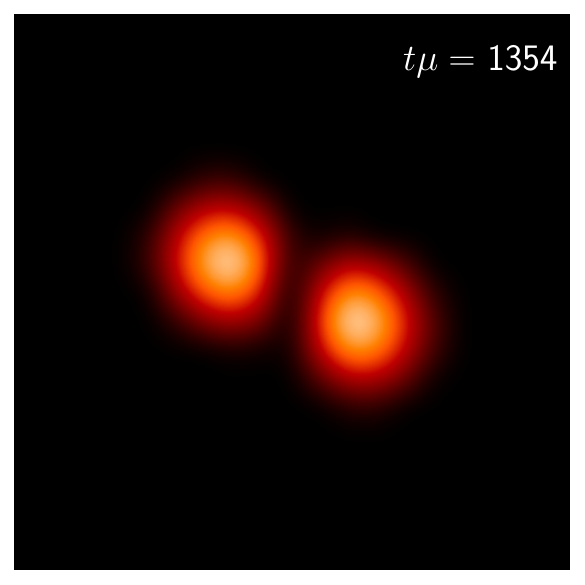}\hspace{-0.1cm}
    \includegraphics[height=0.135\textheight]{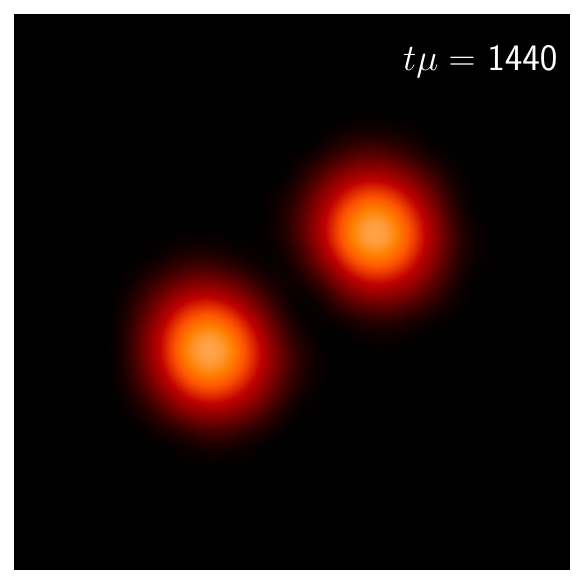}\hspace{-0.1cm}
    \includegraphics[height=0.135\textheight]{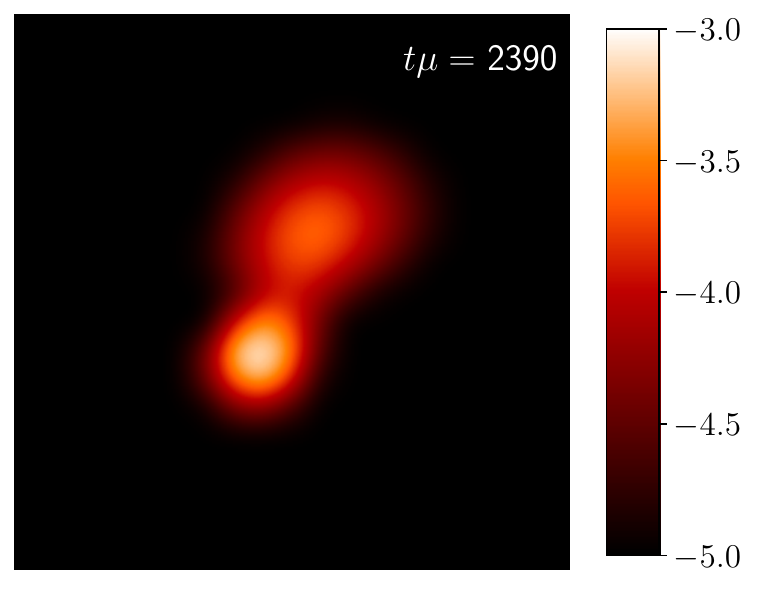}}
    \\
    \subfigure[~~$q=1.5$]{
    \includegraphics[height=0.135\textheight]{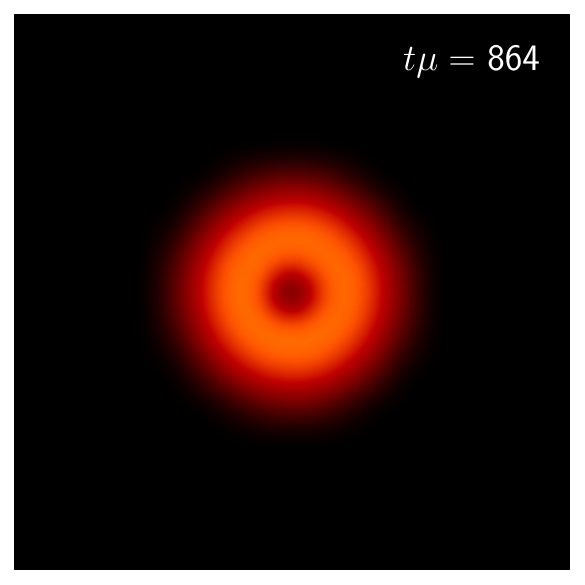}\hspace{-0.1cm}
    \includegraphics[height=0.135\textheight]{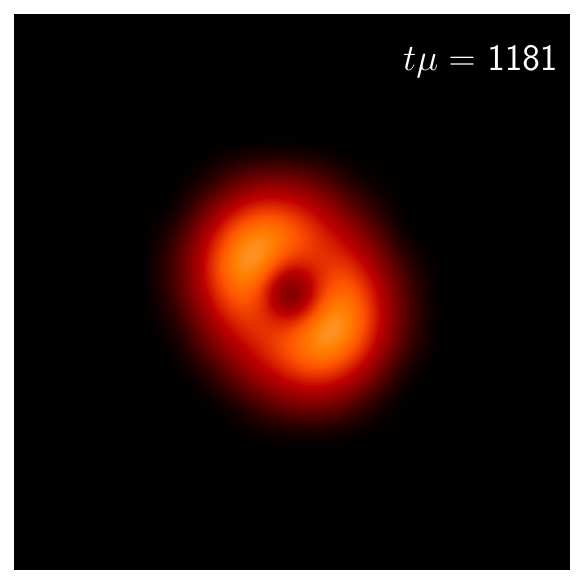}\hspace{-0.1cm}
    \includegraphics[height=0.135\textheight]{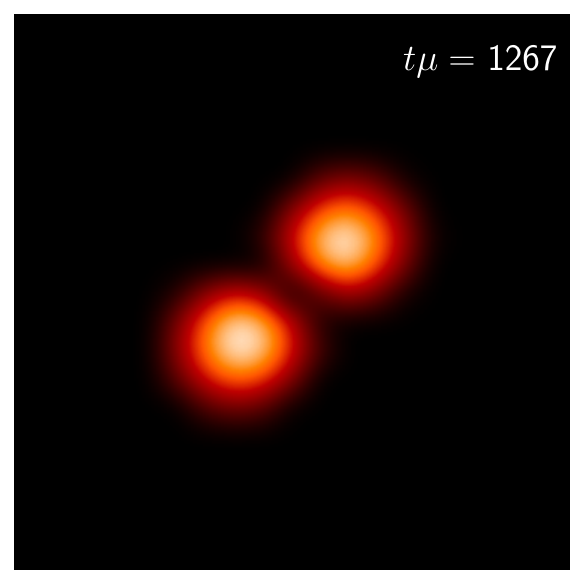}\hspace{-0.1cm}
    \includegraphics[height=0.135\textheight]{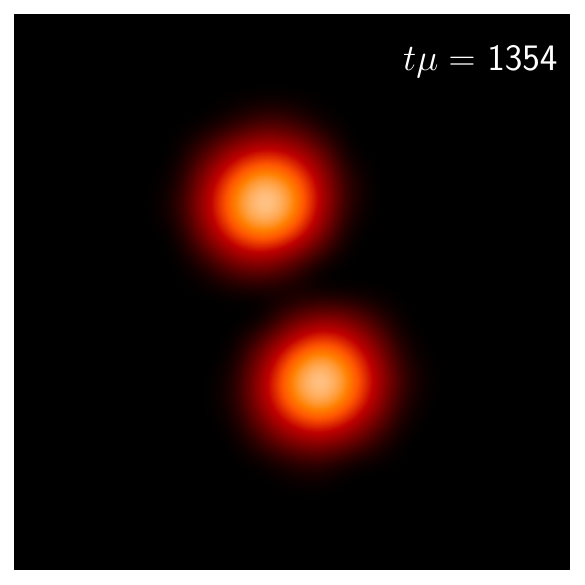}\hspace{-0.1cm}
    \includegraphics[height=0.135\textheight]{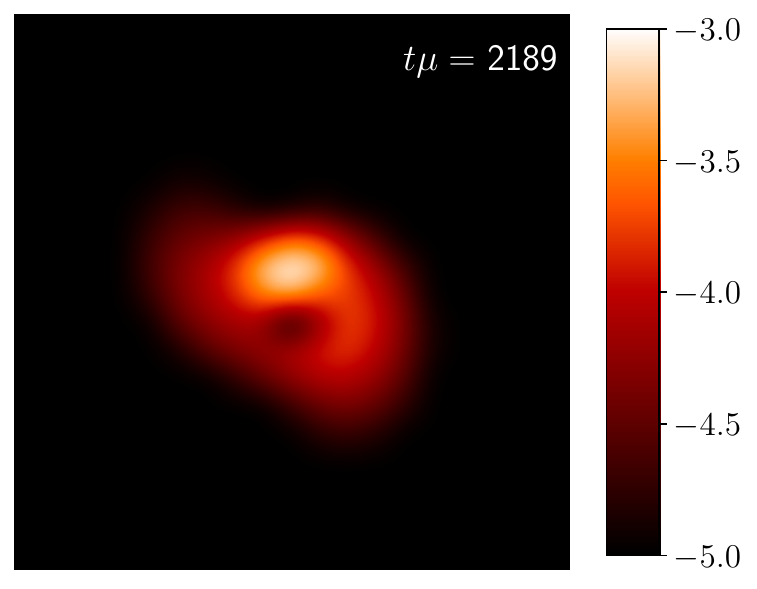}}
    \caption{Snapshots of the magnitude of the scalar field  $|\Phi|$ (log scale) in the equatorial plane at selected times for the rotating ($m=1$) star 
    with $\omega = 0.9\mu$, for both the uncharged (top) and for the  $q=1.5$ (bottom) cases. From left to right: 
    the first column shows the final moment at which the configuration remains stationary. Second column captures the onset of the azimuthal  
    $\tilde{m}=2$ unstable mode. The third and fourth columns depict the two stars as they separate an orbit around each other. The last column shows when the stars merge again, a moment 
    just before collapsing into a black hole. The box side length is $50/\mu$. Notice that the charge $q$ accelerates the entire process.}
    \label{fig:m1_omega0p9_phi}
\end{figure}

Fig.~\ref{fig:m1_omega0p9_phi} displays  
the evolution of the rotating boson star with $\omega = 0.9\mu$, for both the  uncharged and the $q=1.5$ configurations. 
Our results reveal that these two rotating configurations are not only unstable, but their lifetimes shorten as the coupling 
constant $q$ increases. The configurations developed a non-axisymmetric instability, eventually collapsing or transitioning to a spherical state.
Similar behavior is observed in the other $m=1$ configurations displayed in~Table~\ref{tab:electric_table}. They remain stationary during the first  $t\sim 1 \times 10^3/\mu$.
The stars then fragment into two nearly equal-mass spherical stars that orbit each other for a period comparable to the development of the non-axisymmetric instability.
Finally, the fragments  merge, leading to the formation of a black hole shortly after.

The non-axisymmetric instability of spinning boson stars can be qualitatively characterized by a dominant azimuthal mode number,
denoted as $\tilde{m}$ as described in~\cite{DiGiovanni:2020ror,Siemonsen:2020hcg,DiGiovanni:2022mkn}. This integer mode number can be traced during 
the linear phase of the instability by examining the primary mode $e^{i\tilde{m}\varphi}$ in the perturbations of energy density 
or the scalar field amplitude $|\Phi|^2=\Phi\Phi^*$. As the 
linear perturbations enter the nonlinear regime, they manifest as either a monopolar over-density leading to collapse into a transient 
spheroidal configuration for $\tilde{m}=1$, or as a dipolar distribution resulting in the fragmentation of the star into two distinct 
bodies, $\tilde{m}=2$. After these transient phases, the system typically either collapses into a black hole, as the $\omega = 0.9\mu$ cases discussed 
above, or migrates towards a stable spherical configuration, as we will see below. In the specific case shown in Fig.~\ref{fig:m1_omega0p9_phi},
and in all the configurations analyzed in this section, the initial dominant mode is $\tilde{m}=2$.

\begin{figure}
    \includegraphics[width=0.18\textwidth]{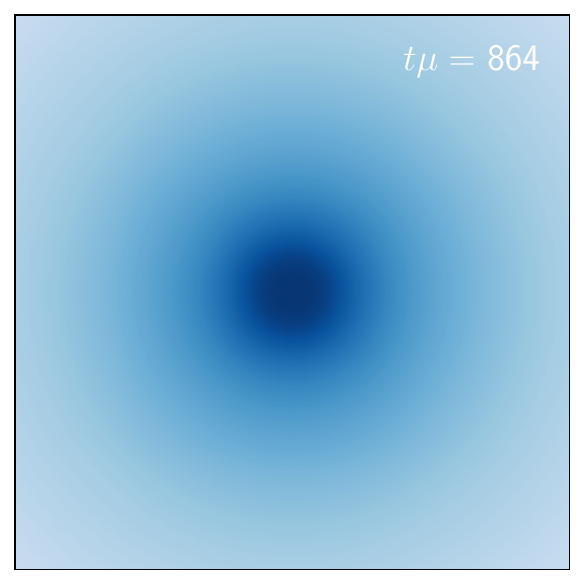}\hspace{-0.1cm}
    \includegraphics[width=0.18\textwidth]{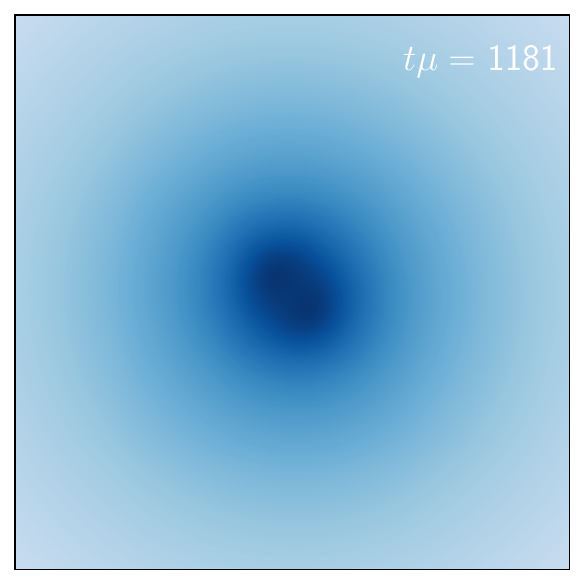}\hspace{-0.1cm}
    \includegraphics[width=0.18\textwidth]{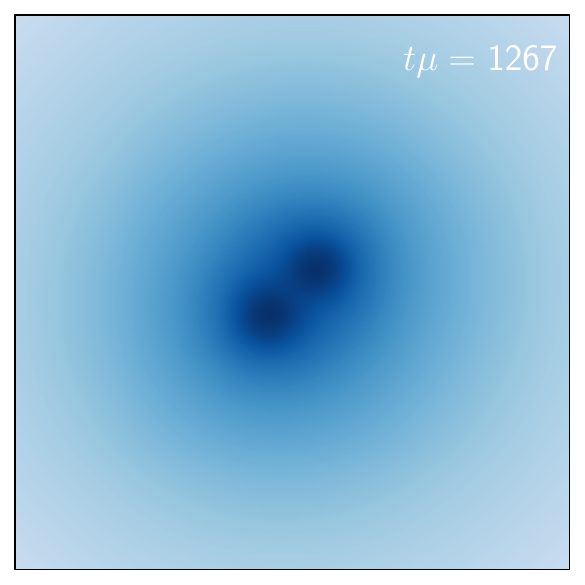}\hspace{-0.1cm}
    \includegraphics[width=0.18\textwidth]{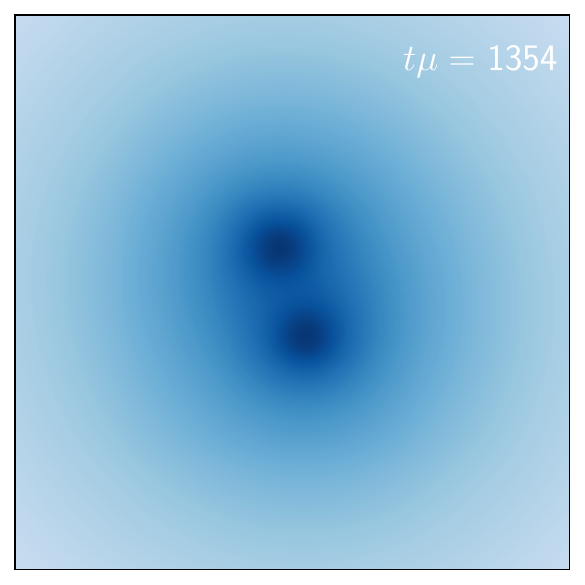}\hspace{-0.1cm}
    \includegraphics[width=0.235\textwidth]{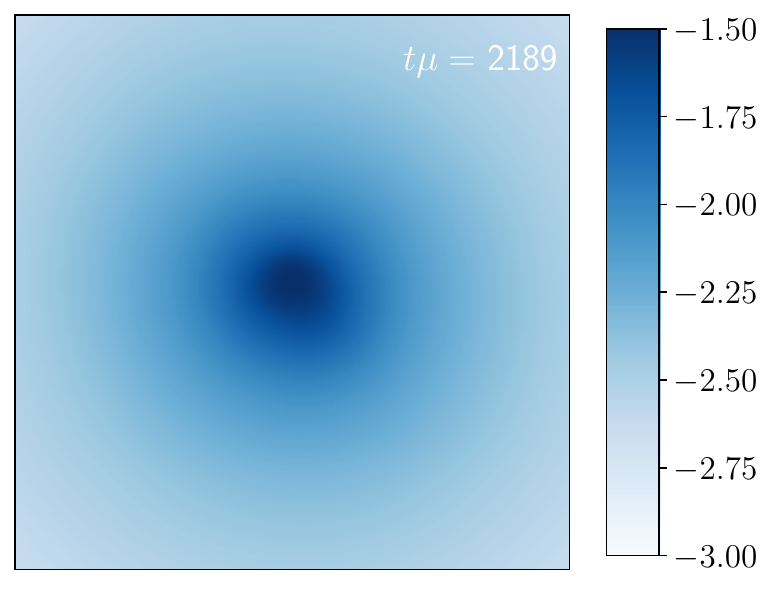}\\
    \vspace{-0.05cm}
    \includegraphics[width=0.18\textwidth]{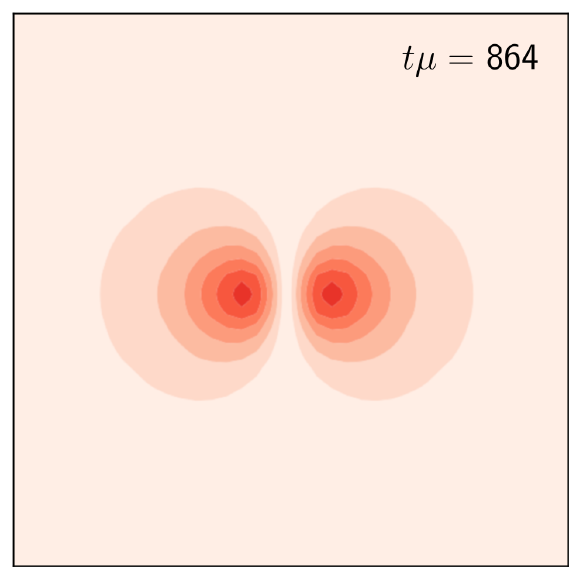}\hspace{-0.1cm}
    \includegraphics[width=0.18\textwidth]{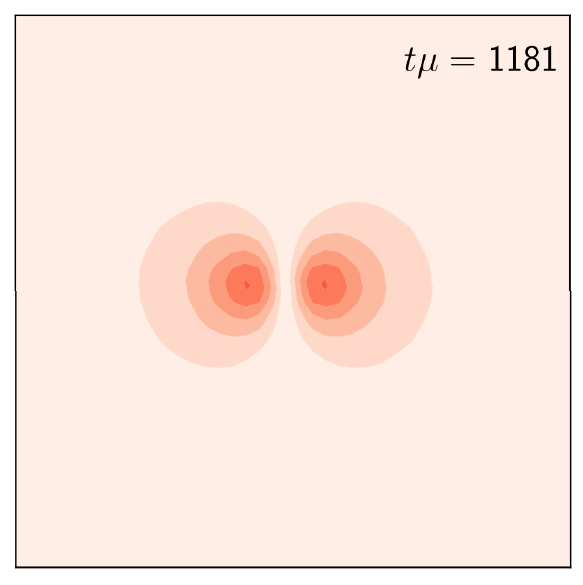}\hspace{-0.1cm}
    \includegraphics[width=0.18\textwidth]{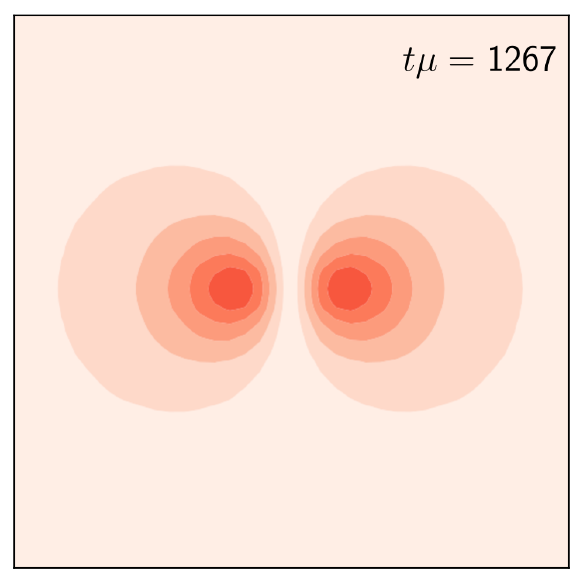}\hspace{-0.1cm}
    \includegraphics[width=0.18\textwidth]{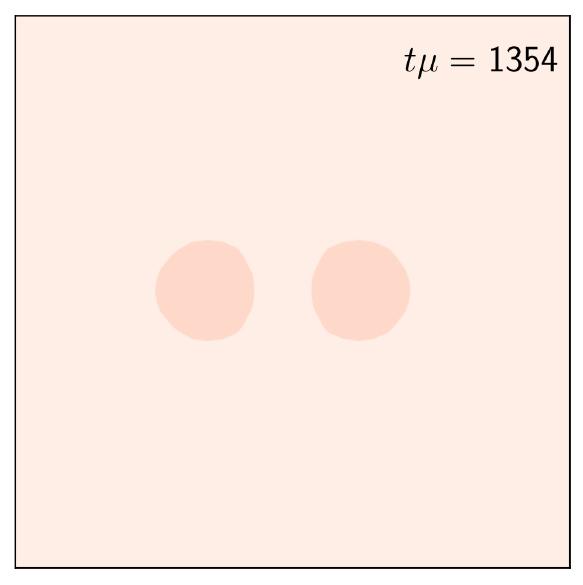}\hspace{-0.1cm}
    \includegraphics[width=0.235\textwidth]{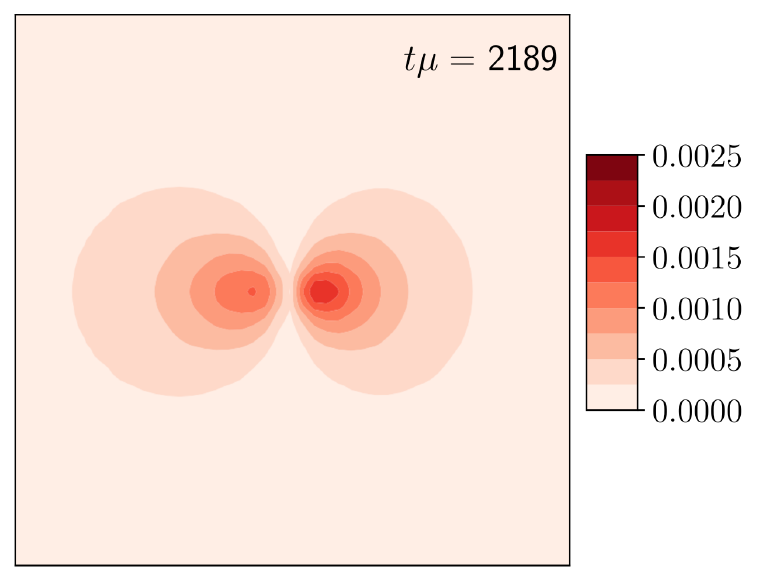}
    \caption{Electromagnetic quantities of the $m=1, \omega=0.9\mu, q=1.5$ boson star configuration, which 
    develops a non-axisymmetric instability and collapses into a black hole at $t=2200/\mu$. The top panel 
    shows the electric potential $\Aphi$ in the equatorial ($z=0$) plane, while the bottom panel displays lines of 
    constant magnetic field in the azimuthal ($y=0$) plane. The box side is $100M$, twice the size
    of Fig.~\ref{fig:m1_omega0p9_phi}.}
    \label{fig:EyB_1T}
\end{figure}

Fig.~\ref{fig:EyB_1T} shows the electromagnetic components during the evolution of a rotating boson star with $\omega = 0.9\mu$
and $q=1.5$ (see bottom panel in Fig.~\ref{fig:m1_omega0p9_phi}). The top panel displays the time component of the electromagnetic 
four-potential in the equatorial ($z=0$) plane, while the bottom panel depicts the lines of constant magnetic field,
\begin{equation}
    B_i = -\frac{1}{2}\epsilon_{i\nu\alpha\beta} n^\nu F^{\alpha \beta}\,,
\end{equation}
in the meridional ($y=0$) plane. From these plots, it is evident that the electric field evolves alongside the fragmentation of the two stars, 
whereas the magnetic field remains in a poloidal configuration throughout the dynamic of the process. In cases where it was possible to follow 
the evolution up to black hole formation, we see that the electric and magnetic fields are no longer around the central object.

\begin{figure}
    \subfigure[~~$q=0$]{
    \includegraphics[height=0.135\textheight]{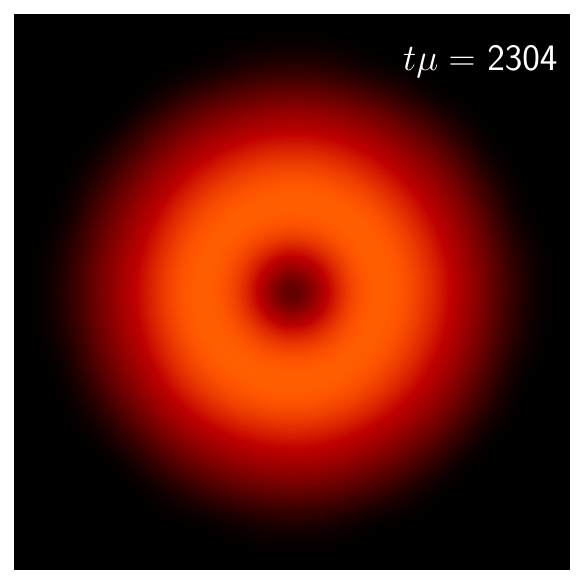}\hspace{-0.1cm}
    \includegraphics[height=0.135\textheight]{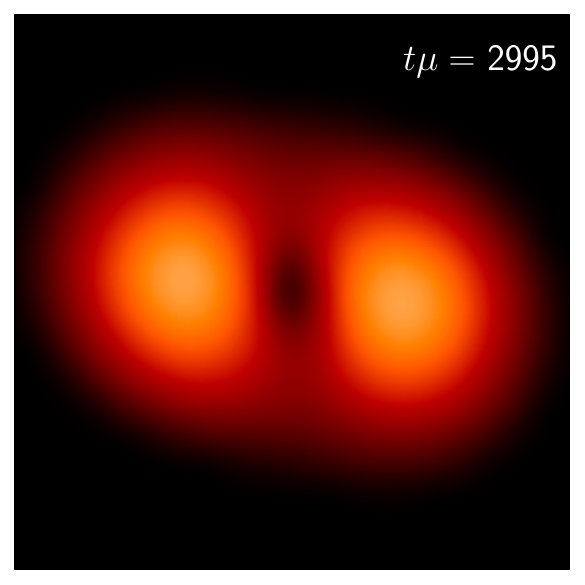}\hspace{-0.1cm}
    \includegraphics[height=0.135\textheight]{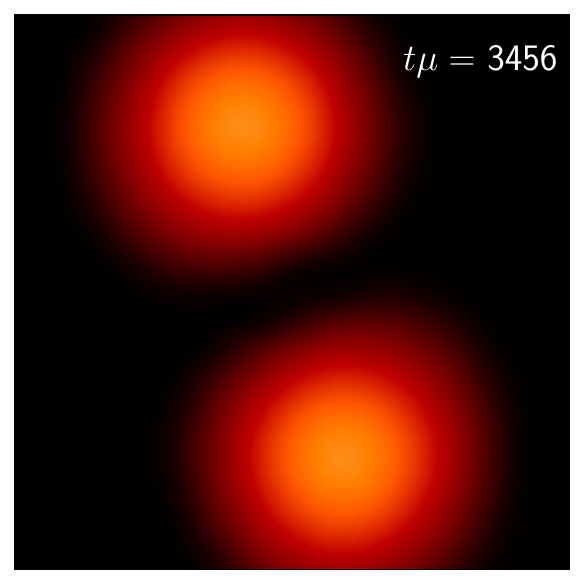}\hspace{-0.1cm}
    \includegraphics[height=0.135\textheight]{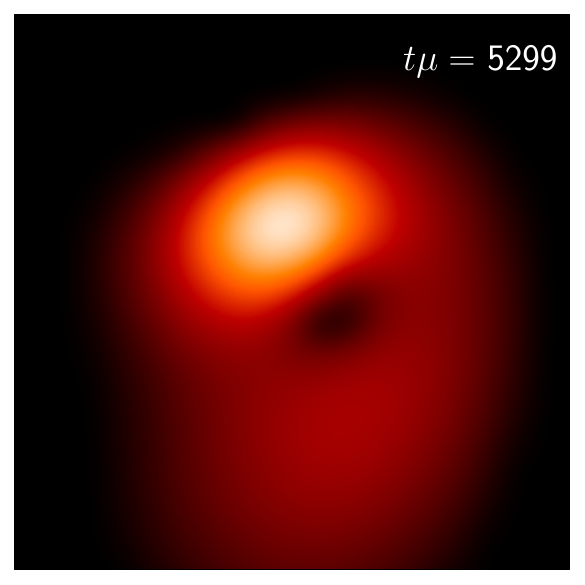}\hspace{-0.1cm}
    \includegraphics[height=0.135\textheight]{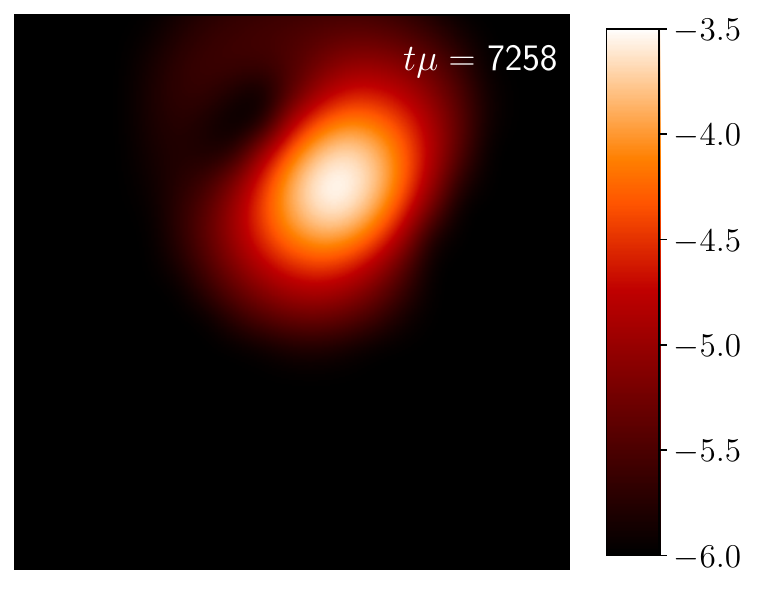}}
    \subfigure[~~$q=2$]{
    \includegraphics[height=0.135\textheight]{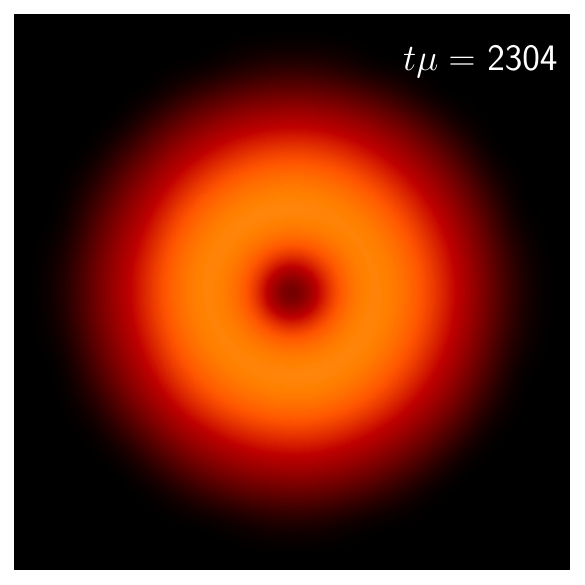}\hspace{-0.1cm}
    \includegraphics[height=0.135\textheight]{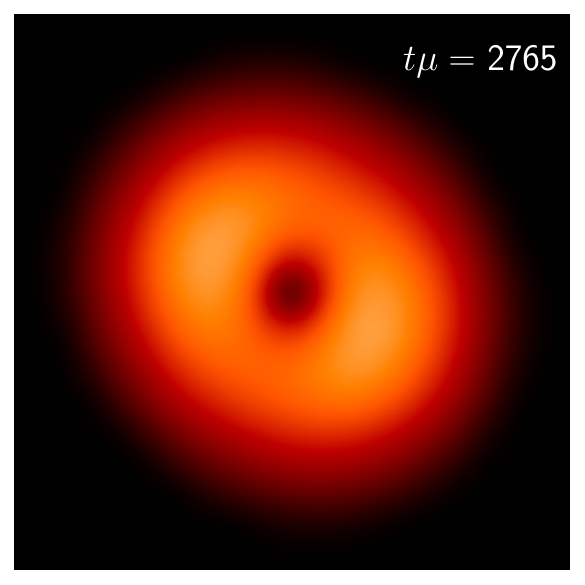}\hspace{-0.1cm}
    \includegraphics[height=0.135\textheight]{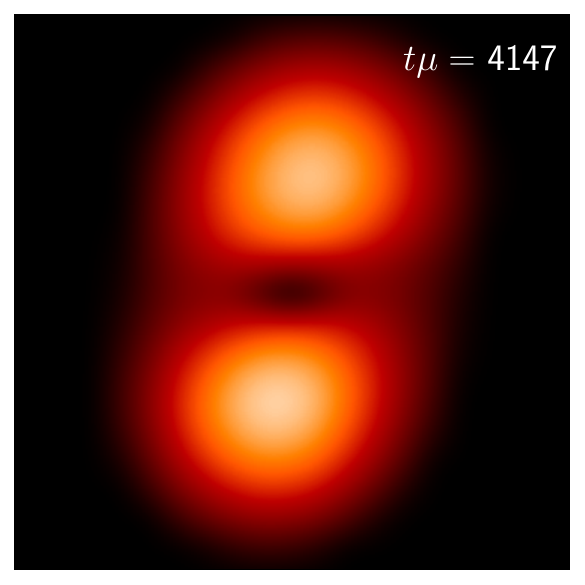}\hspace{-0.1cm}
    \includegraphics[height=0.135\textheight]{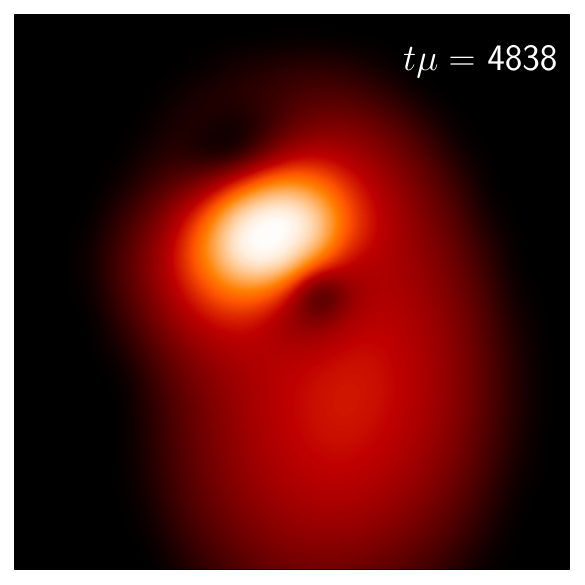}\hspace{-0.1cm}
    \includegraphics[height=0.135\textheight]{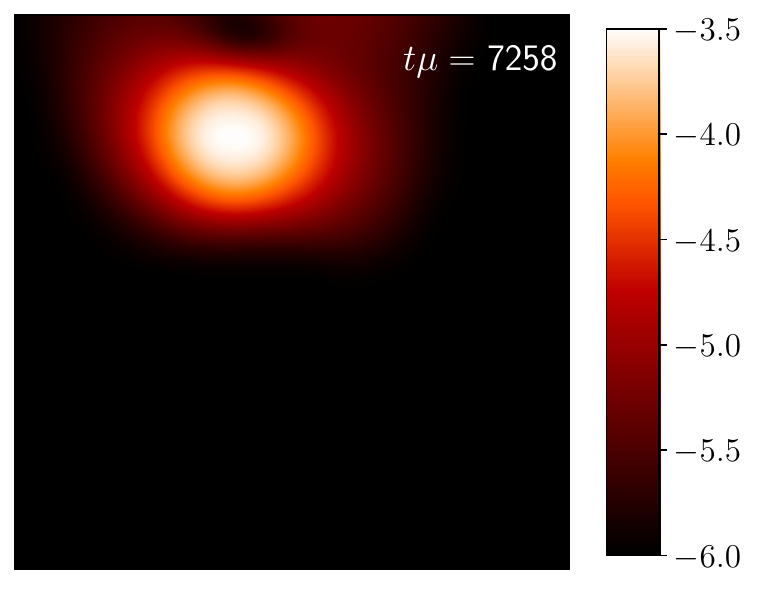}}
    \caption{Snapshots of the scalar field magnitude $|\Phi|$ (log scale) in the equatorial plane at selected times for the rotating ($m=1$) star 
    with $\omega=0.95\mu$, for both the uncharged (top) and for the $q=2$ (bottom) cases. These configurations have a larger radius and are less 
    compact than those in Fig.~\ref{fig:m1_omega0p9_phi}.  Notice that the $\tilde{m}=2$ mode perturbation appears earlier in the charged case. 
    The remnant in both cases is a perturbed spherical charged boson 
    star with some linear momentum. Larger values of the coupling constant $q$, for instance the $q=2.5$ configuration, lead to gravitational 
    collapse to a  black hole. The box side length is $50/\mu$.}
    \label{fig:m1_omega0p95_phi}
\end{figure}

Less massive and compact stars, such as those with $\omega = 0.95\mu$ shown in Fig.~\ref{fig:m1_omega0p95_phi}, do not collapse into a black hole. Instead, after the non-axisymmetric perturbation arises, the fragmented spinning bodies re-collapse into a configuration lacking toroidal morphology, emitting angular momentum and forming a nearly spherical charged boson star. By analyzing the minimum value of the lapse function and the absolute maximum of the scalar field magnitude over time, we observe that for fixed values of $\omega$, the lifetime of the charged rotating stars decreases as the charge $q$ increases. This trend holds for other frequency values as well (as can be seen in Fig.~\ref{fig:m1_omega0p9_phi}). Since increasing the coupling constant $q$ in the first branch of spinning equilibrium configurations leads to an increase in mass, we compared cases with the same mass but different $q$ values. The results clearly show that a higher $q$ also destabilizes the stars.
%

%
\section{Magnetic stars}
\label{sec:two_fields}
We now turn to the problem of studying the stability of electrically neutral magnetized boson stars. Previous studies of these multifield
solutions in the neutral case ($q=0$) have been explored in~\cite{Sanchis-Gual:2021edp}, where it was concluded that 
non-axisymmetric instabilities are developed even in static toroidal stars.
Similar to spinning stars, these static toroidal configurations can be stabilized through superposition with a monopolar boson star, often referred to 
as ``Saturns''~ \cite{Sanchis-Gual:2021edp}. As in the previous sections, we will explore the role of the magnetic field in the dynamics of the system.in \cite{Sanchis-Gual:2021edp}. 
Furthermore, we will demonstrate that self-interactions can indeed stabilize these non-charged configurations, obtaining a stable magnetized self gravitating star.

\begin{figure}
    \includegraphics[width=0.5\textwidth]{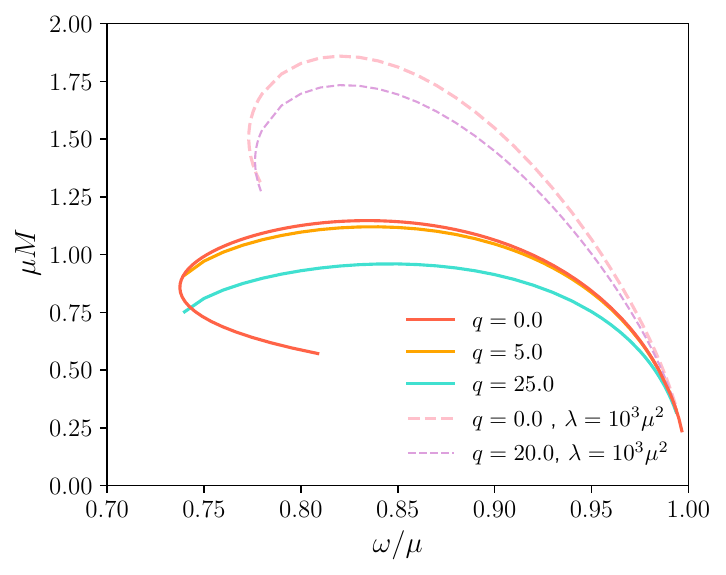}\includegraphics[width=0.49\textwidth]{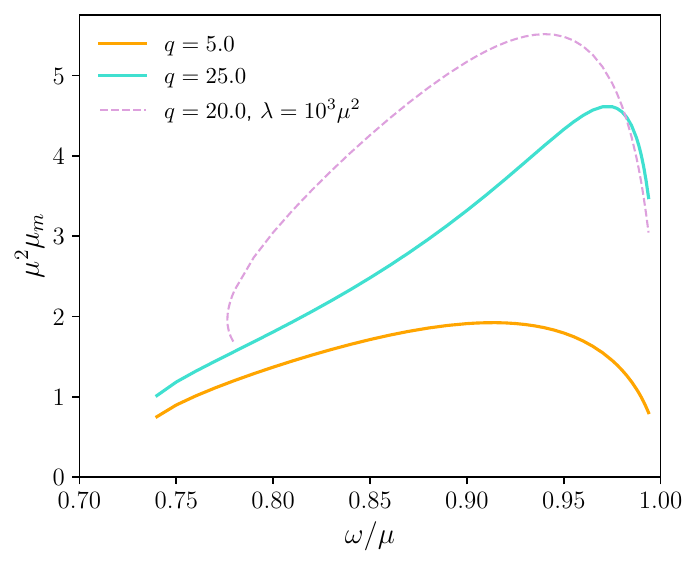}
    \caption{Magnetic boson star with different values of the coupling parameter $q$, shown without (continuous lines) 
    and with (dashed lines) a self-interaction $\lambda$. Left panel shows the
    mass of the configurations {\it vs.} scalar field frequency, while the right panel displays the magnetic dipole moment {\it vs.} scalar field frequency.}
    \label{fig:magBSsequences}
\end{figure}

As discussed in Sec.~\ref{sec:stationary}, the stationary solutions for this case can be obtained by solving the equations presented in the 
Appendix~\ref{app:initial_data_magnetic}. These equations are similar, and even simpler, to those solved for rotating ($m=1$) 
boson stars, allowing us to apply the same  numerical strategies described above. 
Fig.~\ref{fig:magBSsequences} display sequences of magnetic solutions, each parameterized by the coupling constant $q$.  In 
Table~\ref{tab:magnetic_table} we list some of these solutions
along with the relevant classical 
observables.

As shown in Figs.~\ref{fig:BSsequences} and~\ref{fig:magBSsequences}, a key distinction between the rotating charged stars and the static magnetized tori, 
neutral stars is their contrasting mass behavior. For charged stars, the mass increases with the coupling constant $q$, whereas for magnetized 
configurations, the mass decreases. Notably, magnetized configurations do not appear to have an upper limit for $q$.
Configurations with  $q>100$ where previously reported in~\cite{Jaramillo:2022gcq}.
In contrast to the rotating case, in magnetized, neutral stars there is no Coulomb repulsion, so there is no unbounded
interaction that would prevent equilibrium beyond a certain $q$ threshold. 
In fact, the Lorentz force $F_{\mu\alpha}J^\alpha$ derived from the electromagnetic potential generated by the magnetostatic boson stars points inward 
in the interior of the star and outward in the central-exterior part~\cite{Jaramillo:2022gcq}, different to what happens in the case of a spherical star 
which points outwards everywhere (except at the origin where the force is zero). The right panel in Fig.~\ref{fig:magBSsequences} depicts the magnetic 
dipole moment, showing that larger mass configurations does not necessarily correspond to those of the larger magnetic moment. 
This occurs because, for bosonic stars in the first branch of solutions, a larger mass leads to a smaller radius, and thus, the morphology of the solutions 
closer to $\omega = \mu$ correspond to larger tori. All configurations in Table~\ref{tab:magnetic_table} were chosen to belong to the 
first branch of their respective solution families.

\begin{table}
  \centering
  \begin{tabular}{cccccc|c}
    \hline\hline
     $\omega/\mu$& $q$ & $\mu M$     & $\mu^2Q$ & $\max\phi$ & $\mu^2 \mu_m$ & Remnant\\ \hline
     $0.9$       & 0.0 & 1.063       & 1.093    & 0.0234     & 0.0           & Mi  \\
     $0.9$       & 0.5 & 1.063       & 1.093    & 0.0234     & 0.208         & Mi  \\
     $0.9$       & 1.0 & 1.063       & 1.093    & 0.0234     & 0.415         & Mi  \\
     $0.9$       & 5.0 & 1.046       & 1.075    & 0.0237     & 1.910         & BH \\
     $0.9$       & 25.0& 0.914       & 0.937    & 0.0264     & 3.319         & BH \\
     $0.86$      & 0.0 & 1.135       & 1.175    & 0.0347     & 0.0           & BH \\
     $0.86$      & 5.0 & 1.111       & 1.149    & 0.0352     & 1.766         & BH \\
     $0.86$      & 25.0& 0.957       & 0.986    & 0.0398     & 2.632         & BH \\
     $0.9295$    & 0.0 & 0.957       & 0.977    & 0.0159     & 0.0           & Mi  \\
     $0.9295$    & 5.0 & 0.945       & 0.965    & 0.0161     & 1.900         & BH \\
     $0.9295$    & 25.0& 0.838       & 0.855    & 0.0177     & 3.909         & BH \\
    \hline\hline
  \end{tabular}
  \caption{Two fields selected configurations, for several values of the frequency and of the coupling constant; all of them without self interaction term. The column labels are the same as those in Table~\ref{tab:electric_table}, we see that these configurations either migrate, Mi, or collapse to a black hole, BH.
    }
  \label{tab:magnetic_table}
\end{table}
Next, we aim to evolve selected cases from Table~\ref{tab:magnetic_table}. To accomplish this, it is relatively straightforward to generalize the full BSSNOK system \eqref{eq:BSSNfull} in the code. We implement the necessary modifications in the evolution code to handle scenarios where the physical model consists of multiple scalar fields, $\Phi_{(i)}$, each with a charge $q_i$. Specifically, since we are interested in evolving the magnetic boson star system \cite{Jaramillo:2022gcq}, we focus on the case with two scalar fields and charges $q_1 = -q_2 = q$, as described in Eq.~\eqref{eq:action_two}.

There are notable similarities in the dynamical properties of the charged rotating case.  Consistent with previous findings, we observe that 
all free scalar field solutions exhibit a non-axisymmetric instability.
Additionally, the lifetime of these configurations decreases monotonically as the charge $q$ increases, while keeping  $\omega$  fixed. 
However, we have identified significant differences compared to the spinning torus discussed in Sec.~\ref{sec:single_field}.

\begin{figure}
    \subfigure[~~$|\Phi_{+}(t)|^2-\phi^2$]{
    \includegraphics[height=0.115\textheight]{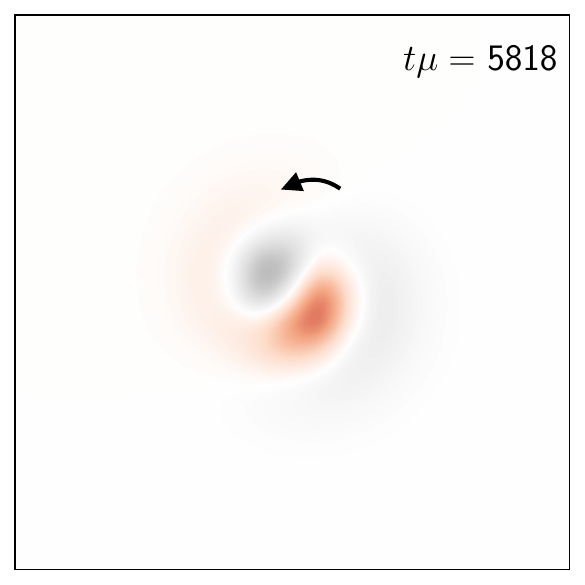}\hspace{-0.1cm}
    \includegraphics[height=0.115\textheight]{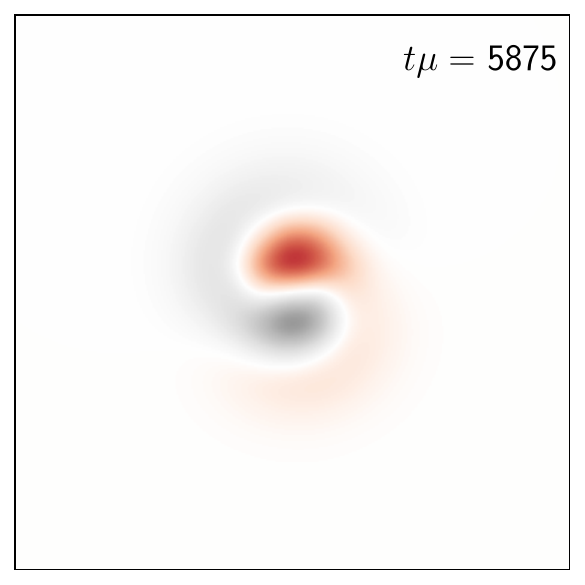}\hspace{-0.1cm}
    \includegraphics[height=0.115\textheight]{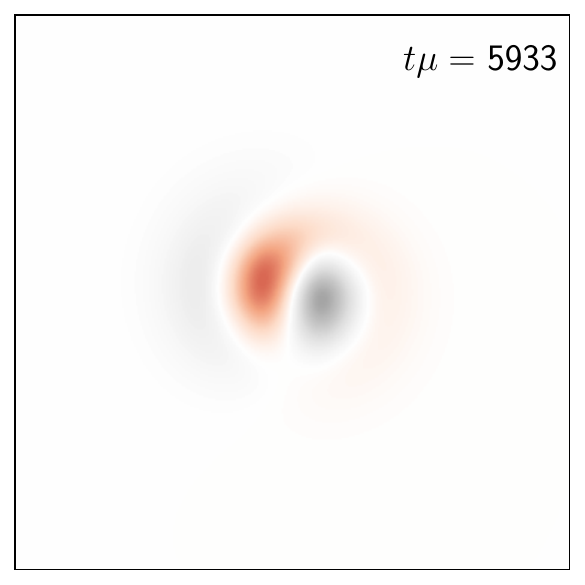}\hspace{-0.1cm}
    \includegraphics[height=0.115\textheight]{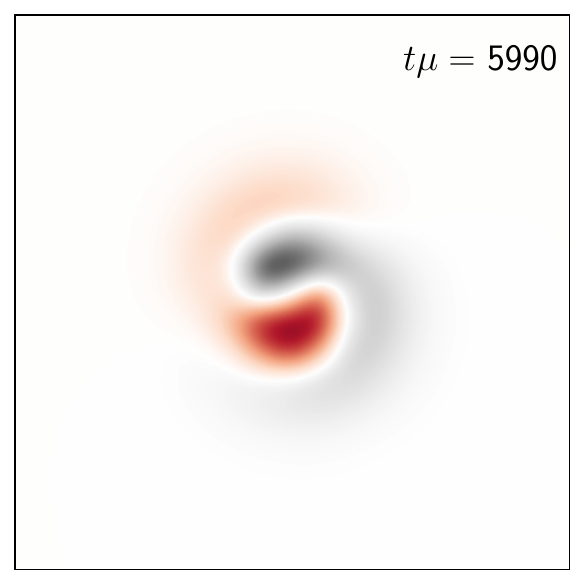}\hspace{-0.1cm}
    \includegraphics[height=0.115\textheight]{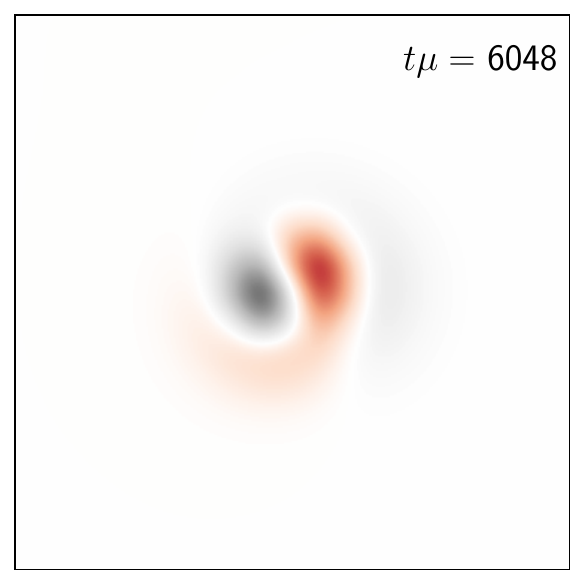}\hspace{-0.1cm}
    \includegraphics[height=0.115\textheight]{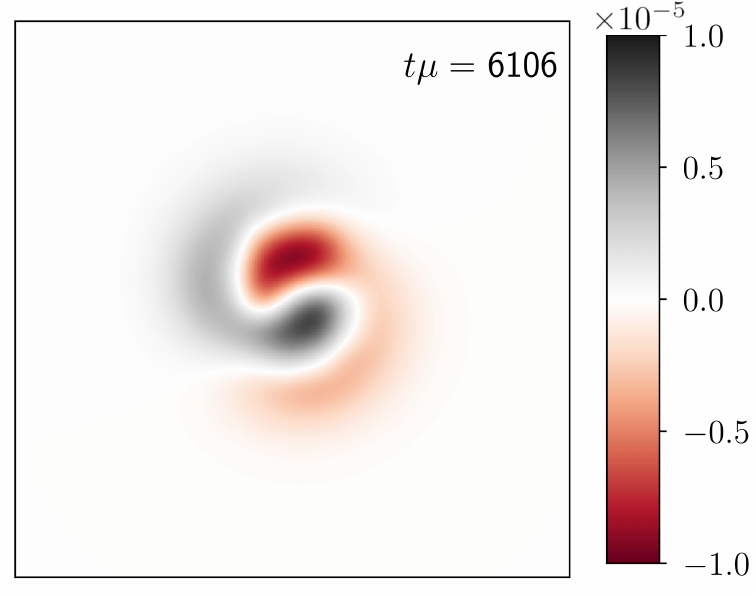}
    }
    \subfigure[~~$|\Phi_{-}(t)|^2-\phi^2$]{
    \includegraphics[height=0.115\textheight]{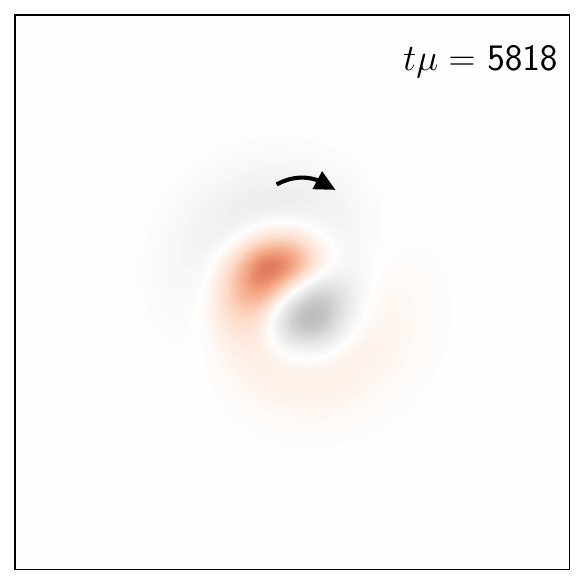}\hspace{-0.1cm}
    \includegraphics[height=0.115\textheight]{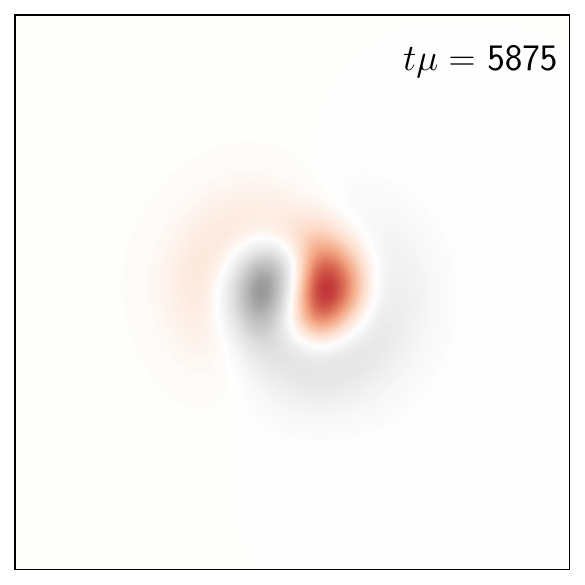}\hspace{-0.1cm}
    \includegraphics[height=0.115\textheight]{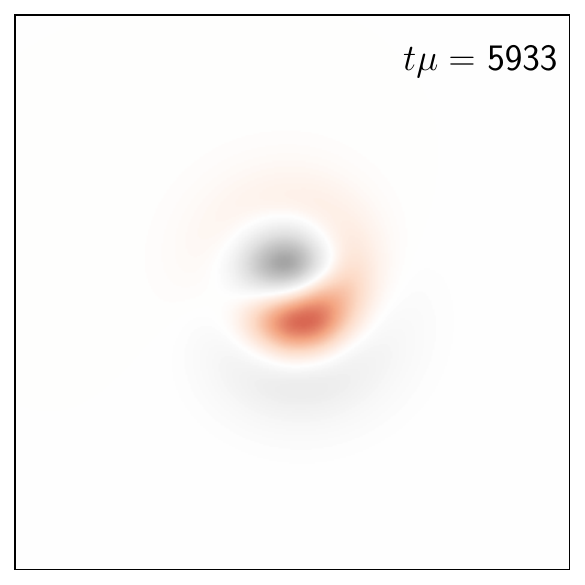}\hspace{-0.1cm}
    \includegraphics[height=0.115\textheight]{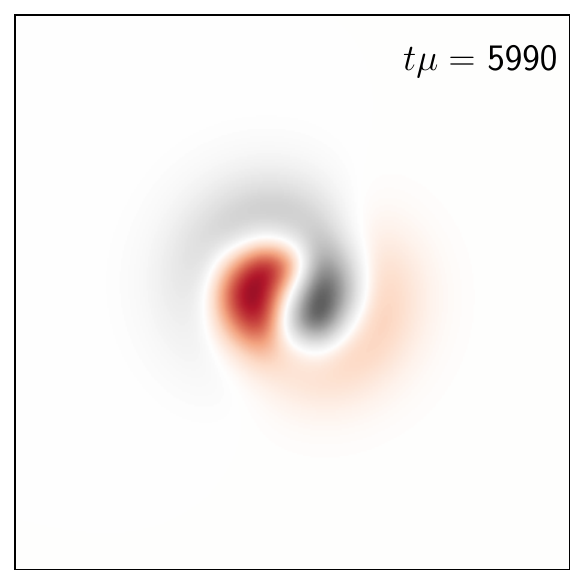}\hspace{-0.1cm}
    \includegraphics[height=0.115\textheight]{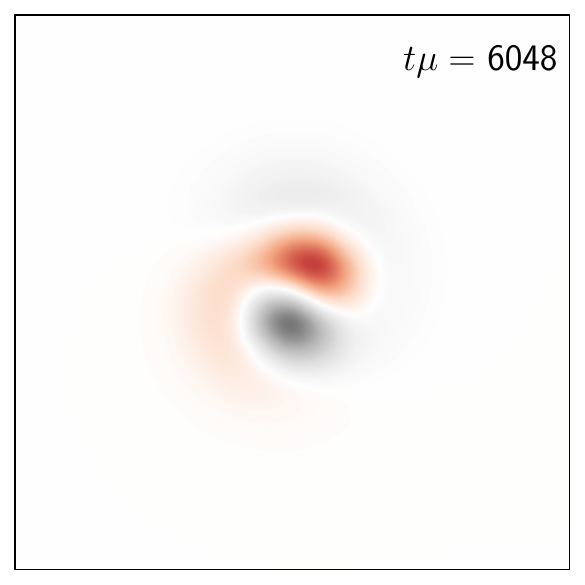}\hspace{-0.1cm}
    \includegraphics[height=0.115\textheight]{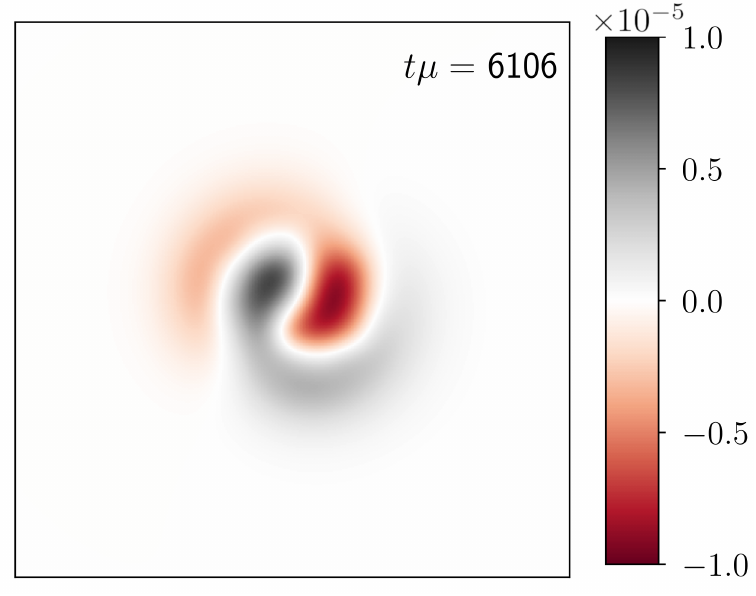}
    }
    \caption{Development of the non-axisymmetric instability for the static toroidal star with $q = 0$. The snapshots show the evolution of the 
    difference in the scalar field magnitudes compared to the initial data $|\Phi_{\pm}(t)|^2 - |\Phi_{\pm}(t=0)|^2 \equiv |\Phi_{\pm}(t)|^2-\phi^2$ 
    in the equatorial plane. Top panel shows the positively charged scalar field, while the bottom panel displays the negatively charged scalar field, 
    both at corresponding time steps. The perturbation exhibits an azimuthal mode $\tilde{m} = 1$. The box side length is $50/\mu$, and the color bar range 
    is $\pm 10^{-5}$. }
    \label{fig:mag_omega0p9_q0}
\end{figure}

We find that all neutral ($q=0$) configurations in~Table~\ref{tab:magnetic_table} develop a $\tilde{m}=1$ instability,
which tends to appear earlier in the more massive cases compared to the less massive ones.
Examining the plots of the two scalar fields, we observe that during the linear stage, the $\tilde{m}=1$ dipole structure (see Fig.~\ref{fig:mag_omega0p9_q0}) of the quantities  $|\Phi_{\pm}(t)|^2 - |\Phi_{\pm}(t=0)|^2 \equiv |\Phi_{\pm}(t)|^2 - \phi^2$  rotates in opposite directions in the  equatorial plane: anti-clockwise for  $\Phi_+$  and clockwise for $\Phi_-$. Upon entering the nonlinear regime of the perturbation, other higher angular azimuthal modes emerge, and both scalar fields form a single monopolar configuration. This configuration may represent a highly perturbed superposition of two concentric spherical boson stars. While we clearly identify the monopolar structure of the remnant, similar to the charge boson star configurations, the scalar field profiles remain far from settling even at very late times, preventing us from identifying a definitive end-state. Therefore, we cannot rule out the possibility of a multi-state (spherical) star \cite{Bernal:2009zy}. The more compact neutral case, with  $\omega = 0.86\mu$ , collapses into a black hole (see Table~\ref{tab:magnetic_table}).

\begin{figure}[H]
    \subfigure[~~$|\Phi_{+}(t)|^2-\phi^2$]{
    \includegraphics[height=0.135\textheight]{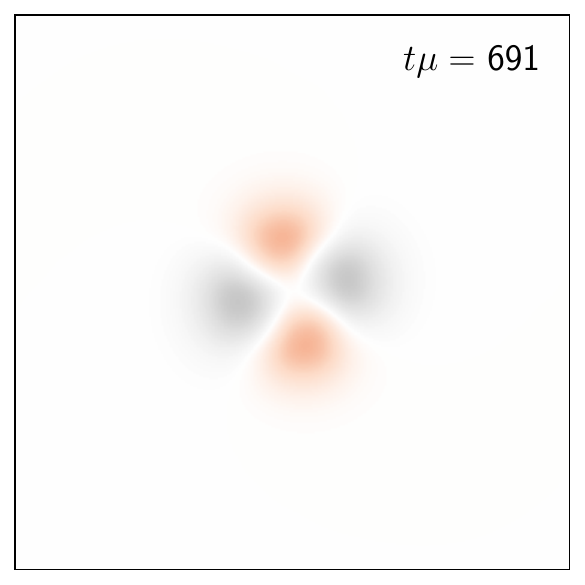}\hspace{0.075cm}}\rule{0.025cm}{0.135\textheight}\subfigure[~~$\log_{10}|\Phi_+(t)|$]{
    \includegraphics[height=0.135\textheight]{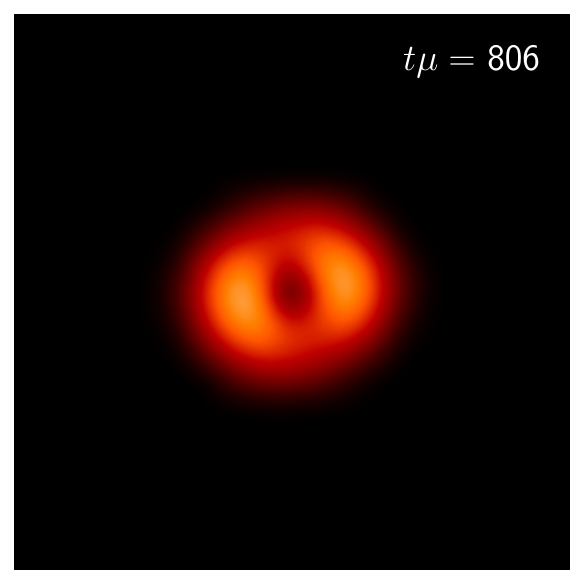}\hspace{-0.2cm}
    \includegraphics[height=0.135\textheight]{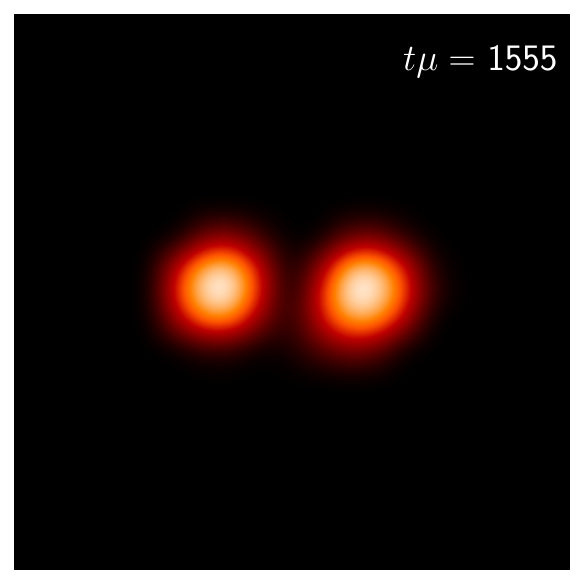}\hspace{-0.2cm}
    \includegraphics[height=0.135\textheight]{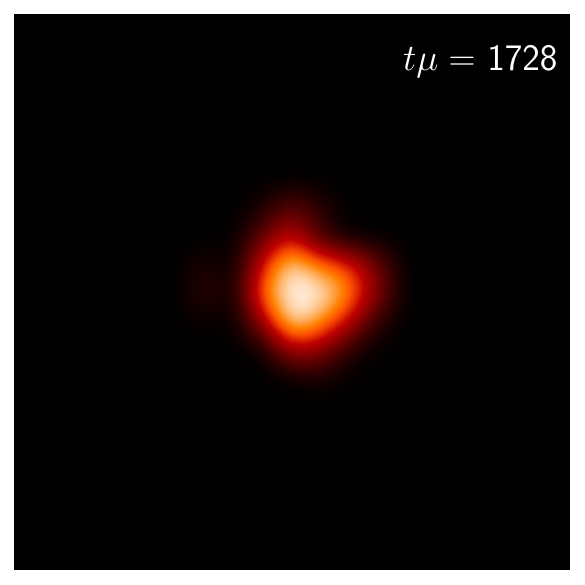}\hspace{-0.2cm}
    \includegraphics[height=0.135\textheight]{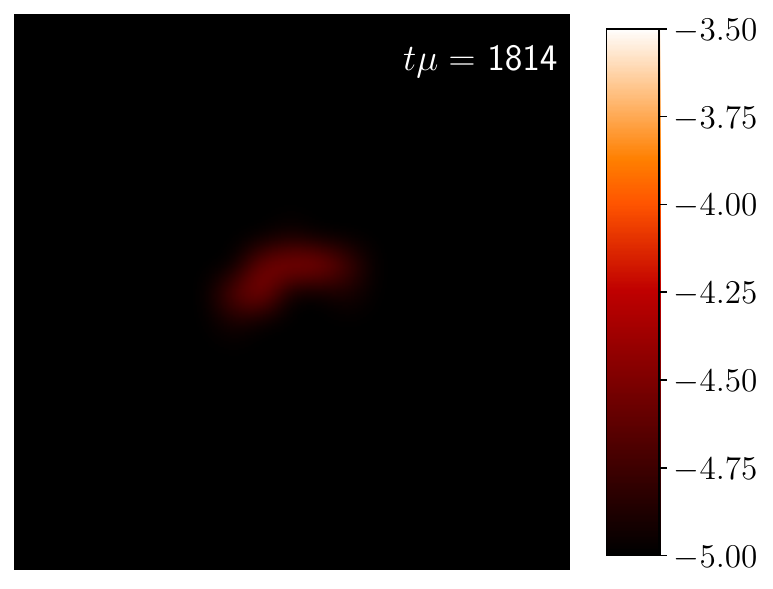}}
    \subfigure[~~$|\Phi_{-}(t)|^2-\phi^2$]{
    \includegraphics[height=0.135\textheight]{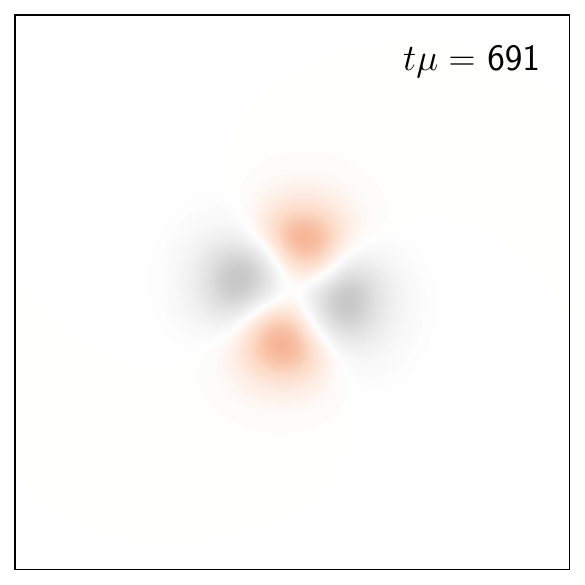}\hspace{0.075cm}}\rule{0.025cm}{0.135\textheight}\subfigure[~~$\log_{10}|\Phi_-(t)|$]{
    \includegraphics[height=0.134\textheight]{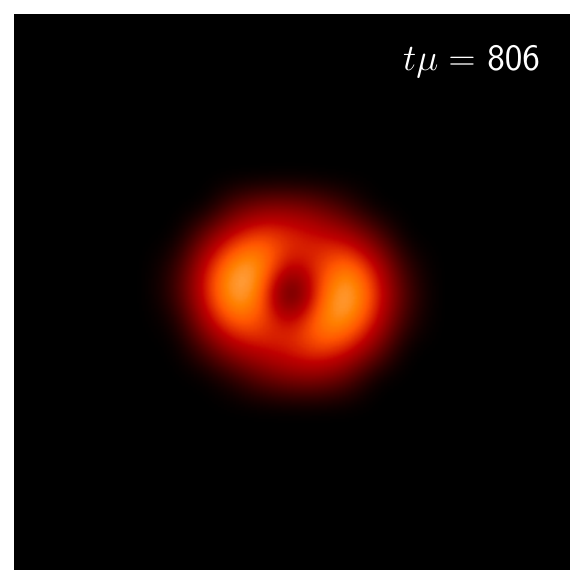}\hspace{-0.2cm}
    \includegraphics[height=0.134\textheight]{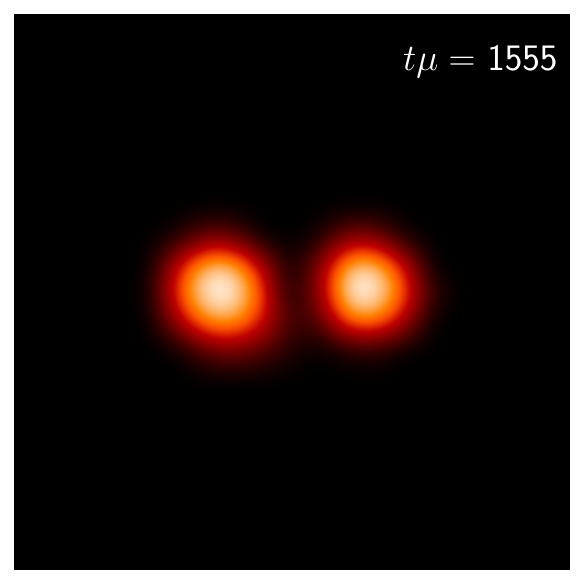}\hspace{-0.2cm}
    \includegraphics[height=0.134\textheight]{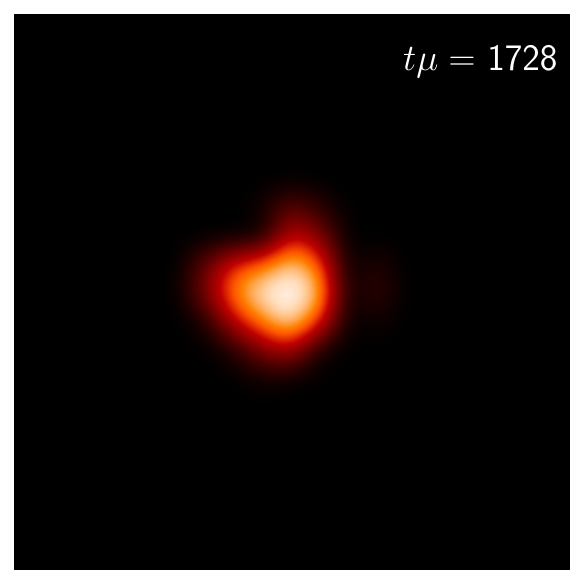}\hspace{-0.2cm}
    \includegraphics[height=0.134\textheight]{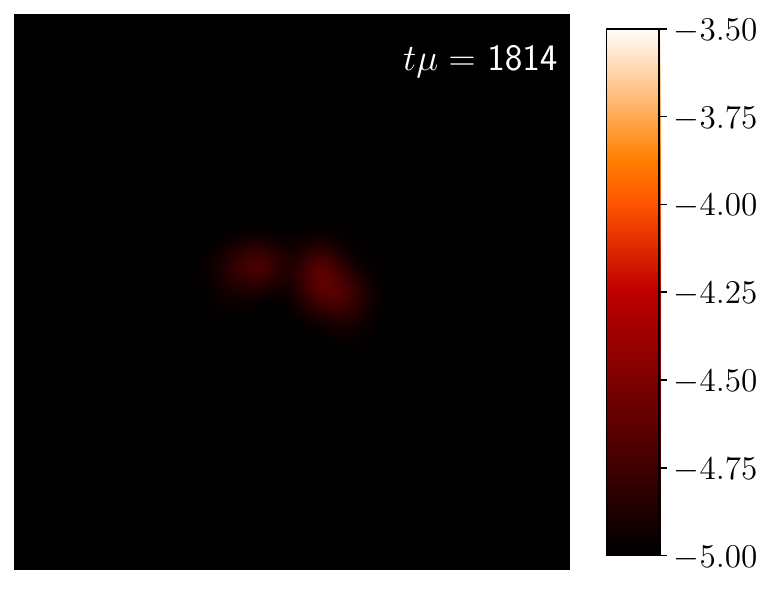}}
    \caption{Magnetic star with $\omega = 0.9\mu$ and $q = 5$ developing a non-axisymmetric $\tilde{m}=2$ instability. Panels (a) and (c) show the difference $|\Phi_{-}(t)|^2-\phi^2$ 
    from the initial data at the moment when the linear instability manifests for the $\Phi_+$ and $\Phi_-$ fields, respectively. Panels (b) and (d) display the corresponding scalar fields on a logarithmic scale.  The first column captures an instant of time where the nonlinear phase is already in 
    operating in the system, followed by a snapshot of the non-rotating dipolar configuration, which persists for approximately $500/\mu$ time units. This is followed by an image of the two lumps merging, and finally, the fourth column illustrates the collapse into a black hole. The color coding for panels (a) and (c), as well as the box size for panels (a) through (d), is consistent with that in Fig.~\ref{fig:mag_omega0p9_q0}.
    }
    \label{fig:phi_mag_omega0p9_q0p5}
\end{figure}

When a non-zero coupling parameter $q$ is considered, we find that the dominant mode of the perturbation changes from $\tilde{m}=1$ to $\tilde{m}=2$. We also examined the cases with smaller $q$ values, specifically  $q = 0.5$  and  $q = 1$ , both of which developed a clear $\tilde{m}=2$ dominant perturbation. However, for the larger case of  $q = 25$, a distinct, albeit brief, linear $\tilde{m}=1$ mode reappears. Similar to the spinning torus cases, the dipolar structure in  $\Phi_\pm$  (which manifests as a quadrupolar pattern in  $|\Phi_{\pm}(t=0)|^2 - |\Phi_{\pm}(t)|^2$; see Fig.~\ref{fig:phi_mag_omega0p9_q0p5}) persists for a period of time, even during the nonlinear perturbation regime, when the differences from the initial configuration become significant.

\begin{figure}[H]
\centering
\includegraphics[width=0.26\textwidth]{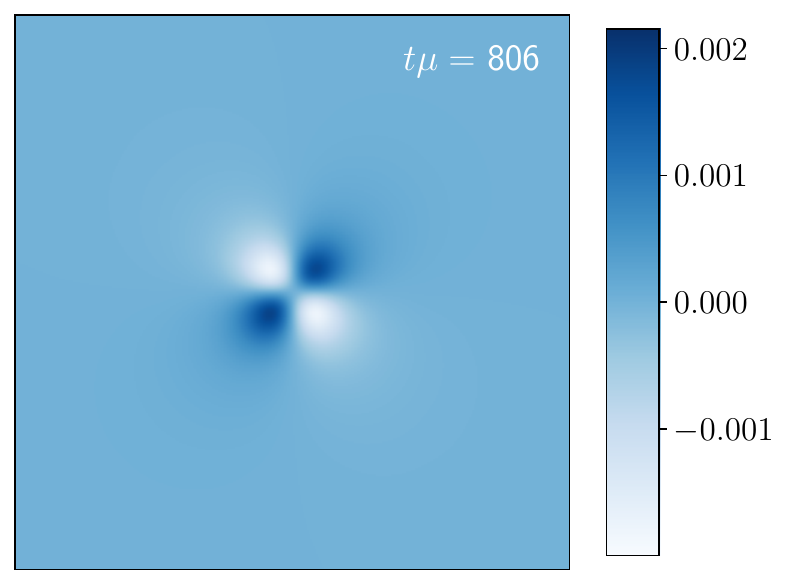}\hspace{-0.1cm}
    \includegraphics[width=0.26\textwidth]{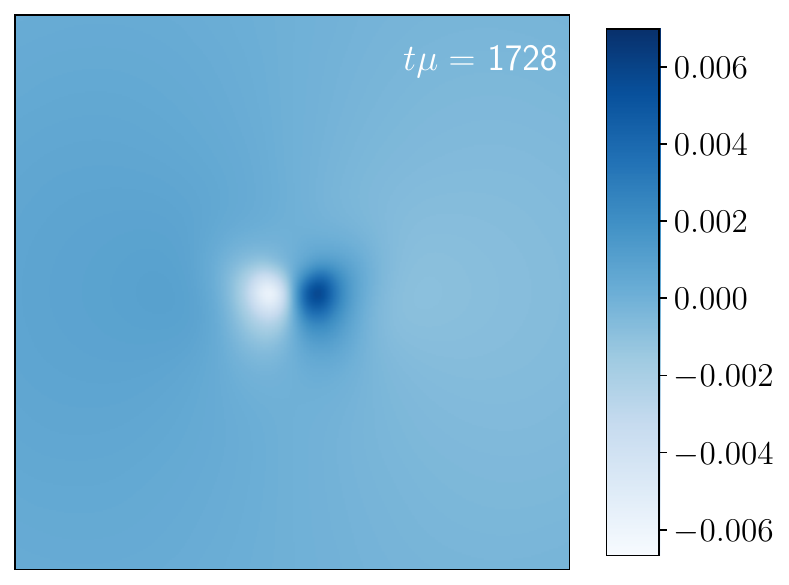}\hspace{-0.1cm}
    \includegraphics[width=0.26\textwidth]{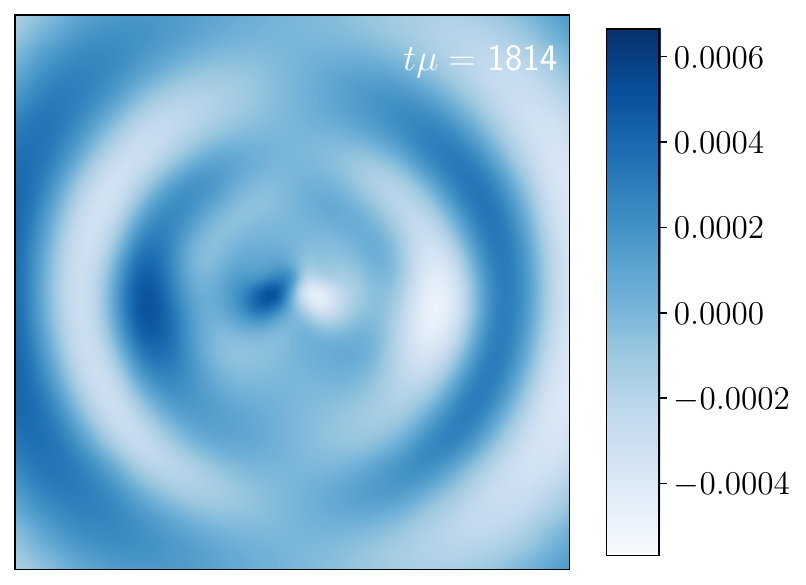}\\
    \vspace{-0.05cm}
    \includegraphics[width=0.26\textwidth]{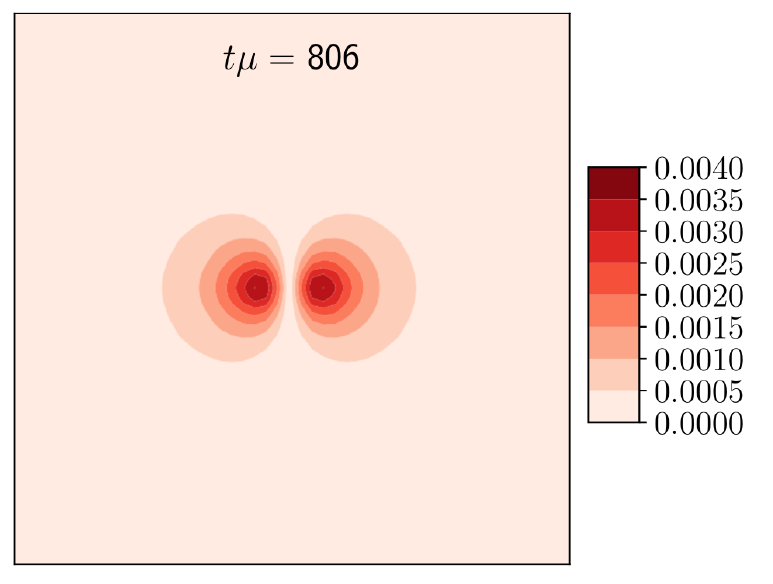}\hspace{-0.1cm}
    \includegraphics[width=0.26\textwidth]{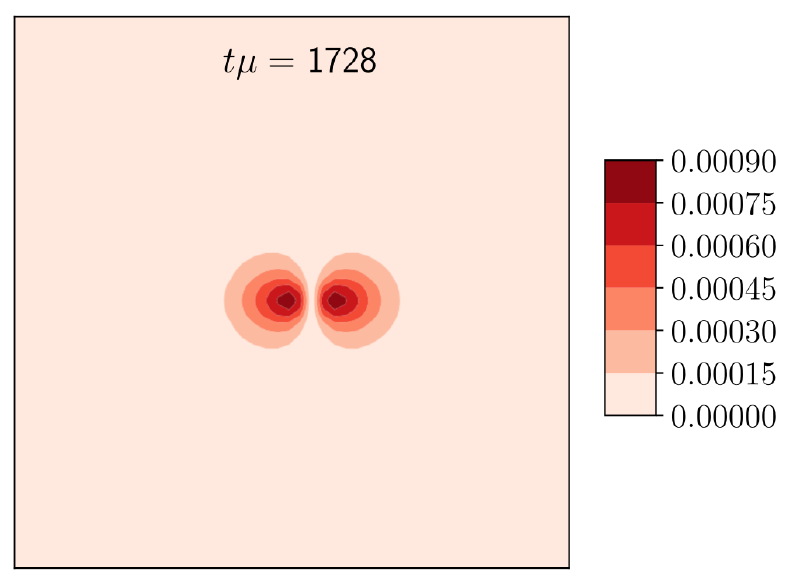}\hspace{-0.1cm}
    \includegraphics[width=0.26\textwidth]{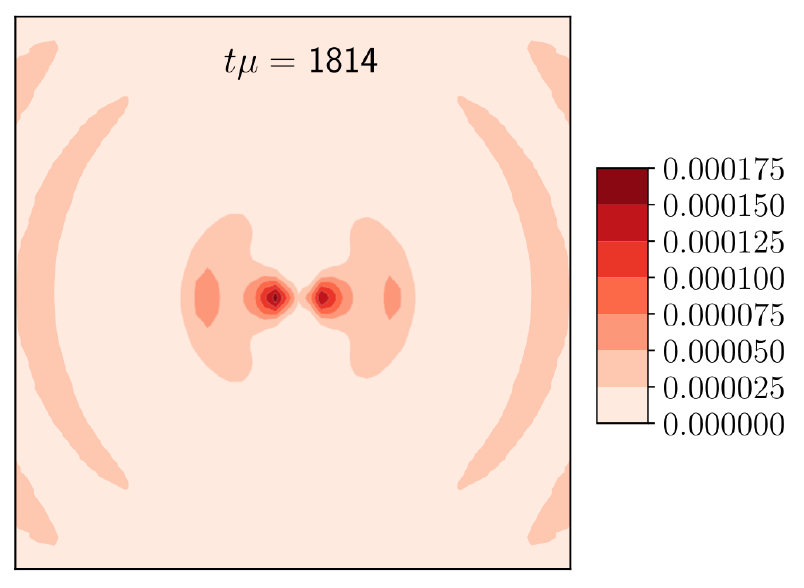}
        \caption{
     Electromagnetic quantities of the boson star configuration  in Fig.~\ref{fig:phi_mag_omega0p9_q0p5}.  The top panel shows
the electric potential $\Aphi$ in the equatorial  ($z=0$) plane, while the  bottom panel displays the  lines of constant magnetic field in the 
azimuth ($y=0$) plane. The side of the box is $100/\mu$.   
Note that the snapshots in this figure correspond to the same time frames as those in columns 1, 3, and 5 of Fig.~\ref{fig:phi_mag_omega0p9_q0p5}. At the initial times the electromagnetic fields remain bounded and then it is seen the emission.}
    \label{fig:mag_omega0p9_q0p5}
\end{figure}

Once the transient dipolar structure formed during the nonlinear stage (for instance the one shown in third column of Fig.~\ref{fig:phi_mag_omega0p9_q0p5}),
we observe the orientation of it remains practically fixed across all cases where it occurs. 
The dipole initially aligns along the symmetric lines  $\varphi = \pi/2 \pm \eta$, where  $\eta$  is a random angle. During this phase, the dynamical magnetic field also exhibits a dipolar morphology (see Fig.~\ref{fig:mag_omega0p9_q0p5}). In contrast, the electric field is quadrupolar during the linear phase but transitions to a dipolar structure afterwards. Following this stage, the two matter lumps merge into a single remnant. For the  $\omega = 0.9 \mu$ configurations with  $q = 0.5$  and  $q = 1$, the final state evolves into a horizonless monopolar structure, while all other configurations collapse into a black hole. Throughout this process, both gravitational and electromagnetic waves are emitted.  

\subsection{Scalar self-interactions. Stable configurations} 
\label{Sec:stable-conf}

Our previous numerical results  
show that the free-field boson stars with $m=1$ are unstable. In what follow, we probe
if  self-interactions can stabilize, or at least extend the lifetime of these systems. As discussed in the Sec.~\ref{sec:stationary} and explored in~\cite{DiGiovanni:2020ror,Siemonsen:2020hcg}, 
these studies considered a similar theory for an uncharged scalar field, but with the scalar potential including not only the mass term as in Eq.~\eqref{eq:Vphi}, but also a positive quartic self-interaction 
term:\footnote{Note that while the Lagrangians in \cite{DiGiovanni:2020ror} and \cite{Siemonsen:2020hcg} differ by a factor of two, the self-interaction terms coincide after appropriately rescaling the scalar 
field conventions to match.}
\begin{equation}\label{eq:quartic_potential}
V(|\Phi|) = \mu^2|\Phi|^2 + \lambda |\Phi|^4.
\end{equation}
The studies in~\cite{Siemonsen:2020hcg} found that in fact the non-axisymmetric instability of rotating boson stars can be suppressed within certain regions of parameter space 
(including also different self-interaction potentials). Specifically, for configurations with $\omega \simeq 0.9\mu$, in \cite{Siemonsen:2020hcg} it was carefully determined that 
a minimum value of $\lambda = 132.2\mu^2$ is sufficient to suppress the instability. 

After discussing the role of the electromagnetic field in the evolution of the rotating boson stars, and observing that the coupling constant (charge) not only fails
to prevent collapse but actually accelerates it, we now examine the effect of the electromagnetic field in stable configurations which include the self-interaction term.
We have confirmed that no signs of instability occur when considering $\lambda = 500\mu^2$ for the uncharged $\omega = 0.9\mu$ case, with stability maintained up to a maximum evolution time of 
$t = 10^4/\mu$. 

\begin{figure}
    \includegraphics[width=0.8\textwidth]{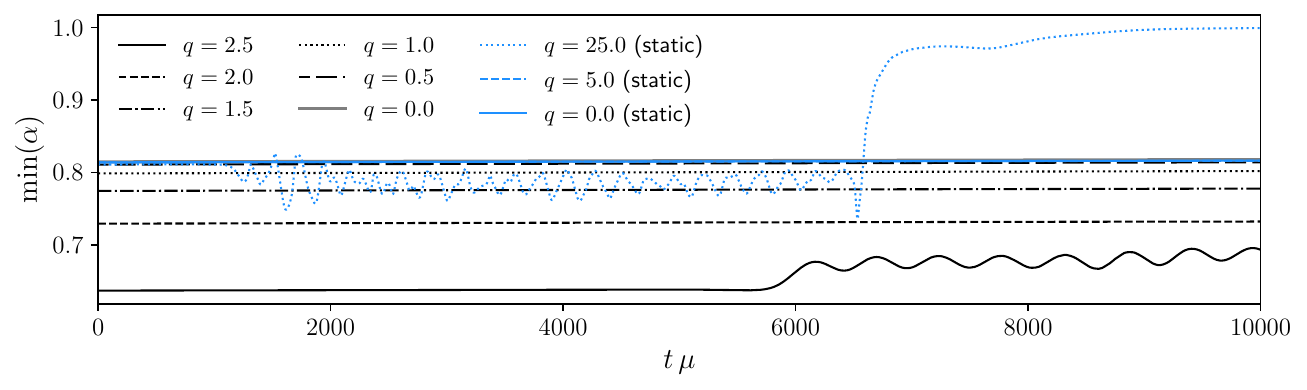}\\
    \hspace{0.2cm}\includegraphics[width=0.8\textwidth]{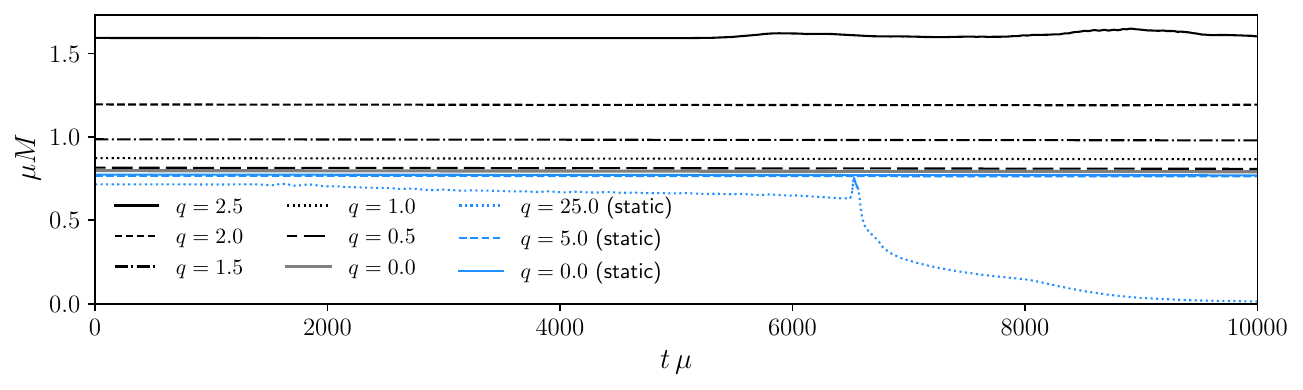}
    \caption{Evolution of spinning and magnetic boson stars with a self-interaction potential (see Eq.~\eqref{eq:quartic_potential}). The 
    configurations shown correspond to $\omega = 0.9\mu$, $\lambda = 133.2\mu^2$, and various values of the coupling constant $q$. 
    The black lines represent spinning boson stars, while the blue lines denote (static) magnetic configurations. The top panel displays 
    the minimum value of the lapse function, and the bottom panel shows the total scalar field energy, calculated using the Komar integral 
    in Eq.~\eqref{eq:komarM}.
    }
    \label{fig:lambda_mass}
\end{figure}

Next, we evolved configurations with a fixed frequency while varying the charge $q$. Our results indicate that configurations with charges in the range $0 \leq q \leq 2$ remain stable throughout the evolution. However, the configuration with $q = 2.5$ exhibits a different behavior, transitioning into a spherical, horizonless soliton during the simulation. To illustrate these findings, we present the evolution of six representative configurations in Fig.~\ref{fig:lambda_mass} (black lines), providing clear support for these conclusions.

Finally, we have considered the physical system of two charged, self-interacting scalar fields. The specific quartic self-interactions we consider involve replacing the two terms  $\mu^2 |\Phi_\pm|^2$  in Eq.~\eqref{eq:action_two} with the potentials  $V(|\Phi_\pm|) = \mu^2 |\Phi_\pm|^2 + \lambda |\Phi_\pm|^4$, analogous to the single field case in Eq.~\eqref{eq:quartic_potential}. Therefore, we assume no direct interaction between the two fields. Two sequences of inital data solutions with $\lambda/\mu^2 = 10^3$, one charged and one uncharged, are presented in Fig.~\ref{fig:magBSsequences}. As in the single-field case, the (positive) self-interaction increases the value of the mass throughout the entire first branch. A similar effect is observed in the magnetic dipole moment. 

We examined the stability of configurations with $\omega = 0.9\mu$, both charged and uncharged, and present the results in the corresponding blue lines of Fig.~\ref{fig:lambda_mass}. For the cases with $q = 0$ and $q = 5$, we observed no signs of instability within the time frame of $t < 10^4/\mu$. However, the configuration with $q = 25$ behaves differently, developing a non-axisymmetric instability. Around $t \sim 2000$, it transitions into a spheroidal configuration and ultimately disperses after $t \sim 6000$.


\section{Conclusions}
\label{sec:conclusions}
To fully unlock the potential of multimessenger astronomy, modeling a broad range of gravitational wave sources is essential. Planned 
sensitivity upgrades to ground-based LIGO/KAGRA/Virgo detectors, along with next-generation detectors like Cosmic Explorer 
and the Einstein Telescope, will significantly expand the observable universe for gravitational wave astronomy~\cite{Corsi:2024vvr}. Additionally, 
improvements in electromagnetic observatories~\cite{Haggard:2017qne,DAvanzo:2018zyz} will enhance our ability to detect and correlate EM 
counterparts with gravitational wave signals, providing a more complete  picture of cosmic events. As these advancements increase both the 
volume of space and the range of frequencies we can observe, the ability to detect unexpected phenomena becomes crucial. 
This makes it essential to investigate sources that have not yet received significant attention, such as magnetized exotic 
compact objects. As a step forward in establishing gauged scalar fields (or boson star) systems as viable  multimessenger 
sources,  we studied their stability properties. 

We began by building charged spherical boson stars. We found that increasing the value of the coupling constant $q$ shifts the maximum mass 
to higher values of $\omega$ (see Fig.~\ref{fig:BSsequences}).  All stable configurations lie within the first branch. We evolved them in the 
fully non-linear regime up  to $t\sim 10^4/\mu$, allowing the initial data to be perturbed only by truncation errors. We observed that, 
as expected from linear perturbation analysis, the objects are stable and the  central value of~$\re(\Phi)$ oscillates as $\max(|\Phi|)\cos(\omega t)$.

Next, we turned our attention to the spinning (axisymmetric) $m=1$ solutions, which are unstable in the neutral charge limit. We constructed sequences of solutions and then analyzed 
particular solutions for the two frequencies, $\omega = 0.9\mu$ and $\omega = 0.95\mu$, exploring charge values from $q=0$ to $q=3.2$. As
in the static case, the mass increases with the coupling constant $q$, and the frequency range narrows. While all configurations within the first 
branch of solutions are gravitationally bound, our simulations indicated that they are unstable. The charge $q$ does not act to prevent 
the collapse, but rather it accelerates it. Therefore, the evolution of charged $m=1$ boson stars goes beyond simple analogies of attractive 
or repulsive forces. We observed that the configurations developed a non-axisymmetric instability, eventually collapsing into a black 
hole after a brief phase of fragmentation into orbiting  spherical stars.

We then explore a peculiar configuration which can be constructed in such a way that the total charge is zero but it has a magnetic field. 
A ``toy model'' for the astrophysical objects where the total charge of the particles in the plasma is zero or negligible, but present 
magnetic fields. This is a solution achieved by superimposing two counter-rotating stars with opposite charge~\cite{Jaramillo:2022gcq,Sanchis-Gual:2021edp}. As in the previous case, 
the configurations turn out to be unstable. We observed that the  larger the magnetic field the shorter the lifetime of the star. 
The configurations exhibit intriguing characteristics, such as an inverse relationship between the coupling constant $q$ and the mass of the star. Additionally, there appears to be no upper limit to the magnitude of $q$. Interestingly, the electromagnetic self-interaction of the scalar field does not result in a dominant repulsive force within the star. Instead, the magnetic fields exert both attractive and repulsive influences. We then proceed to evolve these stars for frequencies $\omega = 0.9\mu$, 0.86, and 0.9295, exploring coupling parameters in the range between $q = 0$ and $q = 25$. We find that these non-rotating configurations develop a non-axisymmetric instability. However, unlike other cases, the configurations do not fragment into two blobs of matter during the initial stage of the instability, except within an intermediate range of small, non-zero values of $q$.

Finally, following previous findings that rotating boson stars with a self-interaction term are stable~\cite{DiGiovanni:2020ror,Siemonsen:2020hcg}, 
we explored the role of the electromagnetic field in their stability. Our results show that the configurations remain stable for at least
$t\sim 10^4/\mu$ across the ranges of charge values explored in the free field boson stars.

In this way, we have provided configurations fully consistent with the Einstein-Maxwell-Klein-Gordon theory which set a background where several 
analysis can be made without introducing extra assumptions, usually made {\it by hand}, allowing us to study the motion of charged particles in
such configurations, as well as to perform a consistent study of the multi messenger emission during collisions or collapse.


\acknowledgments
We thank Carlos Herdeiro and Nicolas Sanchis-Gual for their valuable discussions, which significantly contributed to shaping this work. 
DN, VJ, and MR acknowledge the support received during their visit to the Aveiro Gravitational Gr$@$v  group. Additionally, VJ and DN 
express their gratitude for the hospitality and assistance provided by the Departament d’Astronomia i Astrof\'isica, Universitat de 
Val\`encia in the development of this research.
VJ acknowledges support from the National Key R\&D Program of China under grant No.~2022YFC2204603.
This work was partially supported by the CONACyT Network Project No. 376127 ``Sombras, lentes y ondas gravitatorias generadas por objetos compactos astrof\'\i sicos``, and also by the Center for Research and Development in Mathematics and Applications (CIDMA) through the Portuguese Foundation for Science and Technology (FCT - Fundação para a Ciência e a Tecnologia) through projects: UIDB/04106/2020 (with DOI identifier \url{https://doi.org/10.54499/UIDB/04106/2020}); UIDP/04106/2020 (DOI identifier \url{https://doi.org/10.54499/UIDP/04106/2020});  PTDC/FIS-AST/3041/2020 (DOI identifier \url{http://doi.org/10.54499/PTDC/FIS-AST/3041/2020}); CERN/FIS-PAR/0024/2021 (DOI identifier \url{http://doi.org/10.54499/CERN/FIS-PAR/0024/2021}); 2022.04560.PTDC (DOI identifier \url{https://doi.org/10.54499/2022.04560.PTDC}).
DN acknowledges the sabbatical support given by the Programa de Apoyos para la Superaci\'on del Personal Acad\'emico de la Direcci\'on General de Asuntos del Personal Acad\'emico de la Universidad Nacional Aut\'onoma de M\'exico.
MR acknowledges support by the Generalitat Valenciana  (grants CIDEGENT/2021/046 and Prometeo CIPROM/2022/49), and by the Spanish Agencia Estatal de  Investigaci\'on (grants PID2021-125485NB-C21 funded by MCIN/AEI/10.13039/501100011033,  PRE2019-087617, and ERDF A way of making Europe). 
MZ acknowledges financial through FCT project 2022.00721.CEECIND (DOI identifier \url{https://doi.org/10.54499/2022.00721.CEECIND/CP1720/CT0001}).
Further support has been provided by the EU's Horizon 2020 Research and Innovation (RISE) programme H2020-MSCA-RISE-2017 (FunFiCO-777740) and  by  the  EU  Staff  Exchange  (SE)  programme HORIZON-MSCA-2021-SE-01 (NewFunFiCO-101086251). We acknowledge computational resources and technical support of the Spanish Supercomputing Network through the use of MareNostrum at the Barcelona Supercomputing Center (AECT-2023-1-0006), and also the Navigator Cluster at the LCA in U. Coimbra through projects 2021.09676.CPCA and 2022.15804.CPCA.A2.

\appendix

\section{Re-writing the scalar sector for BSSNOK}\label{sec:app:BSSN}
For completeness, we show here the BSSNOK formulation (see \cite{alcubierre2008introduction} for details) of our evolution equations. This system introduces auxiliary variables and in the case of Eqs.~\eqref{eq:dtgamma}-\eqref{eq:dtZ} and Eqs.~\eqref{eq:dtPhi}-\eqref{eq:dtKphi}, implies a change to the conformal metric with unit determinant, ${\tilde{\gamma}}_{ij}=\chi\,\gamma_{ij}, {\tilde{\gamma}}^{jl}=\frac{1}{\chi}\,\gamma^{jl}$ and a change in 
the covariant derivatives when present. To this end, we note that the (three-dimensional) Christoffel symbols of the conformal metric are related to the ones of $\gamma_{ij}$ through the equation ${}^{(3)}{\Gamma^l}_{ij}={\tilde{\Gamma}^l}_{ij} - \frac{1}{2\,\chi}\left(\delta^l_i\partial_j \chi + \delta^l_j\partial_i\chi - {\tilde{\gamma}}_{ij}{\tilde{\gamma}}^{lm}\partial_m\chi\right)$ (cf.~Eq.~(2.8.14) in~\cite{alcubierre2008introduction}). Using the fact that $\gamma^{ij}\gamma_{ij}=3$ it is possible to obtain that the terms involving divergence of vectors transform as follows:
\begin{equation}
    \gamma^{ij}D_i V_j=\gamma^{ij}\tilde{D}_iV_j-\frac{1}{2\chi}\gamma^{ij}V_i\partial_j\chi \, ,
\end{equation}
here, $\tilde{D}$ is the covariant derivative compatible with $\tilde{\gamma}$. 

The dynamical equations for the gravitational, electromagnetic and scalar fields in the BSSNOK formulation are thus:
\begin{subequations}
\label{eq:BSSNfull}
\begin{eqnarray}
\left( \partial_t -  \mathcal{L}_\beta \right)& \tilde \gamma_{ij} & = 
        - 2 \alpha \tilde A_{ij}\, , \\
\left( \partial_t -  \mathcal{L}_\beta \right)& \chi  & = 
        \frac{2}{3} \alpha \chi K\, , \\
\left( \partial_t -  \mathcal{L}_\beta \right)& K & = 
        [\dots] + 4 \pi \alpha (E + S)\, , \\
\left( \partial_t -  \mathcal{L}_\beta \right)& \tilde A_{ij} & = 
        [\dots] - 8 \pi \alpha \left(
          \chi S_{ij} - \frac{S}{3} \tilde \gamma_{ij}
        \right)\, , \\
\left( \partial_t -  \mathcal{L}_\beta \right)& \tilde \Gamma^i & = 
        [\dots] - 16 \pi \alpha \chi^{-1} P^i\, ,\\
\left(\p_t - \Lie_{\beta} \right)& \A_{i} & = 
- \alpha \chi^{-1} \tilde \gamma_{ij} E^j - \alpha \p_i \Aphi - \Aphi \p_i \alpha \\
\left(\p_t - \Lie_{\beta} \right)& Z & = 
\alpha \p_i E^i - \frac{3}{2} \alpha \chi^{-1} E^i \p_i \chi - \alpha \rho_e - \alpha \kappa Z \\
\left(\p_t - \Lie_{\beta} \right)& \Aphi & = 
  \alpha K \Aphi - \alpha \tilde \gamma^{ij} \chi \p_j \A_i
+ \alpha \chi \A_i \tilde \Gamma^i 
+ \frac{\alpha}{2} \A_i \tilde \gamma^{ij} \p_j \chi
- \chi \tilde \gamma^{ij} \A_i \p_j \alpha
- \alpha Z \\
\left(\p_t - \Lie_{\beta} \right)& E^i & = 
\alpha K E^i - \alpha \chi \tilde \gamma^{ij} \mathcal{J}_j
+ \alpha \chi \tilde \gamma^{ij} \p_j Z
+ \chi^2 \tilde \gamma^{ij} \tilde \gamma^{kl} \p_l \alpha \left( \p_j \A_k - \p_k \A_j \right)
\nonumber \\
 & & \quad +  \alpha \chi^2 \tilde \gamma^{ij} \tilde \gamma^{kl}
\left(\tilde D_k \p_j \A_l - \tilde D_k \p_l \A_{j} \right) 
+ \frac{\alpha}{2} \chi \tilde \gamma^{ij} \tilde \gamma^{kl}
\left( \p_j \A_l \p_k \chi - \p_k \A_j \p_l \chi \right)
\\
  \left(\p_t - \Lie_{\beta} \right)& \Phi & = - 2 \alpha K_\Phi \,, \\
  \left(\p_t - \Lie_{\beta} \right)& K_\Phi &  = \alpha \left( K K_{\Phi} - \frac{1}{2} \chi\tilde{\gamma}^{ij} \tilde{D}_i \partial_j \Phi + \frac{1}{4} \tilde{\gamma}^{ij} \partial_i \Phi \partial_j\chi
                  + \frac{1}{2} \mu^2 \Phi \right)
                 - \frac{1}{2} \chi\tilde{\gamma}^{ij} \partial_i \alpha \partial_j \Phi
               \nonumber \\
                    & & \quad + q^2\frac{\alpha}{2}\left(\chi\tilde{\gamma}^{ij}\A_i\A_j-\Aphi^2\right)\Phi - iq\alpha\left(\chi\tilde{\gamma}^{ij}\A_j\partial_i\Phi-2\Aphi K_\Phi\right)      \, .
\end{eqnarray}
\end{subequations}
In the equations for the trace of the extrinsic curvature, $\tilde{A}_{ij}$ and $\tilde{\Gamma}^i$ the $[\dots]$ denote the standard right-hand side of the BSSNOK equations in the absence of source terms.

\section{Initial data}\label{app:initial_data}

We here describe the stationary solutions used as initial data for this work. It is enough to display the full calculations for the rotating charged boson star only, since the electrostatic solution is a particular case of it and, the equations to solve for the magnetostatic solution can be inferred from the obtained system at the end. To this end, we require to write the explicit Einstein, Klein-Gordon and Maxwell equations, Eqs.~\eqref{eq:einstein}, \eqref{eq:maxwell} and \eqref{eq:kg} respectively, for an equilibrium configuration based in the following ans\"atze for the fields,
\begin{align}\label{eq:qi_metric}
    g_{\mu\nu}dx^\mu dx^\nu & = - e^{2 F_0(r,\theta)} dt^2 + e^{2 F_1(r,\theta)}\left(dr^2 + r^2 d\theta^2\right) + e^{2F_2(r,\theta)}r^2\sin^2\theta\left(d\varphi - W(r,\theta) dt\right)^2 \, , \\
\label{eq:ansatz_AA}
    A_\mu dx^\mu &= V(r,\theta) dt + C(r,\theta)d\varphi \, , \\
\label{eq:ansatz_phi3}
    \Phi &= \frac{\phi(r,\theta)}{\sqrt{2}} e^{i(\omega t - m\varphi)} \, .
\end{align}
where the $\sqrt{2}$ is added in order to coincide with the equations presented in \cite{Jaramillo:2022gcq} which differs from the action \eqref{eq:action} by a factor of two.

\subsection{Rotating charged stars}\label{app:initial_data_rotating}

In the quasi-isotropic coordinates chosen for Eq.~\eqref{eq:qi_metric} to arrange the obtained Einstein equations, leading to the following very compact system of partial differential equations,
\begin{align}
\label{eq:eq_N}
\Delta_3 F_0 & = 4\pi A^2(\rho + S)
	+ \frac{B^2 r^2\sin^2 \theta}{2N^2} \, \partial W\partial W
	- \partial F_0 \partial(F_0 + F_2) \\
\label{eq:eq_omeg}
  \tilde \Delta_3 (W r \sin\theta)
	& = - 16\pi \frac{N A^2}{B^2} \frac{P_\varphi}{r\sin\theta}
	+ r\sin\theta  \, \partial W \partial(F_0 - 3 F_2) \\
\label{eq:eq_NB}
   \Delta_2 \left[ (NB-1) r\sin\theta \right]
	& = 8\pi N A^2 B r\sin\theta (S^r_{\ \, r} + S^\theta_{\ \, \theta} ) \\
\label{eq:eq_lnApnu}
  \Delta_2 (F_1 + F_0) & = 8\pi A^2 S^\varphi_{\ \, \varphi} 
 + \frac{3 B^2 r^2\sin^2 \theta}{4 N^2} \, \partial W\partial W
	- \partial F_0  \partial F_0  , 
\end{align}
where the following abbreviations have been introduced $N = e^{F_0}$, $A=e^{F_1}$, $B=e^{F_2}$ and
\begin{eqnarray}
	&  & \Delta_2 := \partial_r^2 + \frac{1}{r}\partial_r
	+ \frac{1}{r^2}\partial_\theta^2 \, , \\
	& & \Delta_3 := \partial_r^2 + \frac{2}{r}\partial_r
	+ \frac{1}{r^2}\partial_\theta^2 + \frac{1}{r^2\tan\theta} \partial_\theta \, , \\
 	& & \tilde\Delta_3 := \Delta_3 - \frac{1}{r^2\sin^2\theta} \, , \\
    & & \partial f_1 \partial f_2 = \partial_r f_1 \partial_r f_2 + \frac{1}{r^2} \partial_\theta f_1 \partial_\theta f_2 \, .
\end{eqnarray}

The source terms are obtained according to their definition in Eq.~\eqref{eq:source}, leading to the following combinations
\begin{eqnarray}
    \rho + S &=& \frac{2}{N^2}\varrho^2\phi^2 - \mu^2\phi^2 + \frac{\vartheta}{A^2} \, ,\\
    P_\varphi &=& \frac{1}{NA^2}\left\{A^2\varrho(qC-m)\phi^2 -W\partial C\partial C -\partial C \partial V \right\} \, ,\\
    {S^r}_r + {S^\theta}_\theta &=& \frac{1}{N^2}\varrho^2\,\phi^2 - \frac{(qC-m)^2}{B^2r^2\sin^2\theta}\phi^2 - \mu^2\phi^2 \, ,\\
    {S^\varphi}_\varphi &=& \frac{1}{2 N^2}\varrho^2\,\phi^2 - \frac{\mu^2\phi^2}{2} + \frac{\left(q C-m\right)^2}{2\,B^2r^2\sin^2\theta}\,\phi^2 + \frac{1}{2 A^2}\left(\vartheta - \partial\phi\partial\phi\right)
\end{eqnarray}
where
\begin{equation}
    \varrho = mW - \omega - q(CW + V) \qquad \text{and} \qquad  \vartheta = \frac{\partial C\partial C}{B^2r^2\sin^2\theta} + \frac{1}{N^2}\left(W^2\partial C\partial C + 2W\partial C \partial V + \partial V\partial V\right) \, .
\end{equation}

The Maxwell equations reduce to the following two,
\begin{equation}
    \Delta_3 V = \partial V\partial(F_0-F_2) - W\left[\Delta_3 C - \partial C\partial(F_0-F_2)\right] - \partial C\partial W - q A^2\varrho\,\phi^2 \, ,
\end{equation}
and
\begin{equation}
\begin{split}
    (2\Delta_2 - \Delta_3) C = &-\partial C\partial(F_0 - F_2) + q A^2 B^2 r^2\sin^2\theta  \left(\frac{W}{N^2}\varrho + \frac{qC-m}{B^2 r^2\sin^2\theta}\right)\phi^2 \\ 
    &+ \frac{WB^2r^2\sin^2\theta}{N^2}\left\{W\left[\Delta_3C-\partial C\partial(F_0-F_2)+\frac{2\partial C\partial W}{W}\right] +\Delta_3 V-\partial V\partial(F_0-F_2) + \frac{\partial V\partial W}{W}\right\} \, ,
\end{split}
\end{equation}
Combining these two equations we obtain equations with second order operators for $V$ and $C$ only:
\begin{align}\label{eq:finaleq_V}
    \Delta_3 V & = \partial V\partial(F_0-F_2) + 2\,W\,\partial C\partial(F_0-F_2) - \partial C \partial W - 2W (\Delta_3-\Delta_2) C - W\Omega + q A^2\left(q V + \omega\right)\phi^2 \, , \\
    \left(2\,\Delta_2 - \Delta_3\right) C & = -\partial C \partial(F_0-F_2) +  \Omega + q A^2\left(qC - m\right)\phi^2 \, ,
\end{align}
with 
$\Omega = B^2r^2\sin^2\theta(W\partial C\partial W + \partial V\partial W)/N^2$. Note that although there is an apparently second order term in the right-hand side of Eq.~\eqref{eq:finaleq_V}, it is actually first order since $\Delta_3 - \Delta_2 = \frac{1}{r}\partial_r + \frac{1}{r^2\,\tan\theta}\,\partial_\theta$.

Finally, the obtained Klein-Gordon equation is
\begin{equation}\label{eq:Klein}
    \Delta_3  \phi = A^2\left(\mu^2-\frac{\varrho^2}{N^2}\right)\phi - \partial\phi\partial(F_0+F_2) + \frac{A^2}{B^2}\frac{(qC-m)^2}{r^2\sin^2\theta}\phi \, .
\end{equation}
To solve the system of partial differential equations \eqref{eq:eq_N}-\eqref{eq:eq_lnApnu}, \eqref{eq:finaleq_V}-\eqref{eq:Klein} we employ the infrastructure and procedure detailed in Sec.~\ref{sec:Numerical methods}.

We translate the initial data generated in this way and in spherical coordinates $x^{i'} = (r,\theta,\varphi)$ to the Cartesian coordinates $x^i=(x,y,z)$, used in the evolution. For this we notice that the metric \eqref{eq:qi_metric} is related to the 3+1 line element 
$ds^2 = -\alpha^2 dt^2 + \gamma_{i'j'}(\beta^{i'}dt + dx^{i'})(\beta^{j'} dt + dx^{j'})$ through $\alpha = N$, $\gamma_{i'j'}={\rm diag}(A^2,A^2r^2,B^2r^2\sin^2\theta)$ and $\beta^{i'} = (0,0,-W)$. Then, the components of the intrinsic metric and shift vector in the Cartesian coordinate base can be calculated using the Jacobian matrices as usual. In particular we obtain $\beta^i = (yW,-xW,0)$. We need to provide $K_{ij}$ to the evolution code and for this we first notice that the only nonzero components in spherical coordinates are $K_{A\varphi} = K_{\varphi A} = -B^2 r^2\sin^2\theta \partial_A W /(2N)$, with $A=r,\theta$.

We also need to provide the fields $\Aphi$, $\A_i$ and $E^i$ at $t=0$. The electromagnetic potential components can be calculated directly from Cartesian components of the gauge field: $A_\mu = (V,-y/(x^2+y^2)\,C,x/(x^2+y^2)\,C,0)$ 
and the definitions below Eq.~\eqref{eq:DefAJ}, recalling that for the line element given by Eq.~\eqref{eq:qi_metric}, the normal vector to the hypersurfaces takes the form $n_\mu=(-e^{F0}, \vec{0})$, $n^\mu=e^{-F0}\,(1,0,0,W)$, we obtain
\begin{equation}
    \Aphi = -\frac{1}{N}(V+CW) \, , \quad \A_i = A_i  \, .
\end{equation}
The electric field can be calculated using Eq.~\eqref{eq:DefEB} first in spherical coordinates, $E^r = (\partial_r V + W\partial_r C)/(NA^2)$, $E^\theta = (\partial_\theta V + W\partial_\theta C)/(r^2NA^2)$, $E^\varphi=0$ and transform, obtaining
\begin{equation}
    \frac{E^x}{x} = \frac{E^y}{y} = \frac{E^r}{r} + \frac{z}{\sqrt{x^2+y^2}} E^\theta \,,\quad \frac{E^z}{z} = \frac{E^r}{r} - \frac{\sqrt{x^2+y^2}}{z} E^\theta \, .
\end{equation}

\subsection{Electrostatic stars}\label{app:initial_data_spherical}
The charged (electrostatic) solutions are a special case of \ref{app:initial_data_rotating} when considering a non-rotating scalar field, $m=0$. The symmetry of the problem then implies that the metric coefficients can be considered as functions of $r$ alone and satisfying the following (in the isotropic gauge),
\begin{equation}
    F_1(r)=F_2(r) \,, ~~ W = 0 \, .
\end{equation}
Then we do not need to solve Eqs.~\eqref{eq:eq_omeg} nor~\eqref{eq:eq_NB} and the remaining Einstein equations become ordinary although coupled differential equations. 

For the electromagnetic potential we can consider
\begin{equation}
    V = V(r) \,, ~~ C = 0 \, .
\end{equation}
and similarly for the scalar field radial profile which will become $\phi = \phi(r)$. In this case only the $E^r$ component of the electric field is different from zero and the extrinsic curvature is zero.

\subsection{Magnetostatic stars}\label{app:initial_data_magnetic}

As explained in Sec.~\ref{sec:stationary}, the magnetic non-spinning configurations arise from a different situation than the electrostatic and charged rotating case, since these are solutions of the two scalar field theory in Eq.~\eqref{eq:action_two}. To obtain the zero angular momentum, zero electric charge particular case, the scalar fields are considered to have opposite charges $q$ and $-q$ and to be rotating in opposite directions. So, considering the ansatz
\begin{equation}
    \Phi_\pm = \frac{\phi(r,\theta)}{2} e^{i(\omega t \mp m\varphi)} \, .
\end{equation}
for the scalar field (notice the convenient factor $\sqrt{2}$, compared with Eq.~\eqref{eq:ansatz_phi3}). In this case the electric component of $A_\mu$ becomes a constant and the ansatz for the scalar field becomes as the one in Eq.~\eqref{eq:ansatz_AA} but with $V = 0$ and the metric simplifies to the case $W=0$. Then it can be shown that the required system to solve consist in the same Einstein equations ~\eqref{eq:eq_N}, ~\eqref{eq:eq_NB}, \eqref{eq:eq_lnApnu} with the same expressions for the sources (but considering $W=0$ and $V=0$ while the Klein-Gordon equation is exactly the same as Eq.~\eqref{eq:Klein} for the single radial profile $\phi(r)$. Finally the modification to the magnetic potential is also small change, leading to the following equation:
\begin{equation}
    \left(2\,\Delta_2 - \Delta_3\right) C = -\partial C \partial(F_0-F_2) + 2\, q A^2\left(qC - m\right)\phi^2 \, .
\end{equation}


\bibliography{ref}
\end{document}